\documentclass[12pt]{article}
\usepackage{amssymb}
\usepackage{amsmath}
\usepackage{accents}
\usepackage{mathrsfs}
\newcommand{\beq}{\begin{eqnarray}}
\newcommand{\eeq}{\end{eqnarray}}
\newcommand{\beqst}{\begin{eqnarray*}}
\newcommand{\eeqst}{\end{eqnarray*}}

\newcommand{\R}{\mathbb R}
\newcommand{\C}{\mathbb C}
\newcommand{\N}{\mathbb N}

\newtheorem{theorem}{Theorem}

\newtheorem{lemma}{Lemma}
\newtheorem{corollary}{Corollary}

\title{Inverse problems for a generalized fractional diffusion equation with unknown history}

\author{Jaan Janno\footnote{Department of Cybernetics. Tallinn University of Technology. Ehitajate tee 5, 19086 Tallinn, Estonia. E-mail: jaan.janno@taltech.ee}}

\date{}

\begin{document}

\maketitle

\abstract{Inverse problems for a diffusion equation containing a generalized fractional derivative are studied. The equation holds in a time interval $(0,T)$ and it is assumed that a  state $u$ (solution of diffusion equation) and a source $f$ are  known  for  $t\in (t_0,T)$ where $t_0$ is some number in $(0,T)$. Provided that
$f$ satisfies certain restrictions, it is proved that product of a kernel of the derivative with an elliptic operator as well as the history of  $f$ for  $t\in (0,t_0)$ are uniquely recovered. 
In case of less restrictions on $f$ the uniqueness of the kernel and the history of $f$  is shown. Moreover, in a case when  a functional of $u$ for  $t\in (t_0,T)$ is  given  the uniqueness of the kernel is proved
under unknown history of $f$. }


\section{Introduction}

In many cases a relative irregularity in a direct problem increases informativeness in an inverse problem. For example,  the smoother the kernel of a Volterra equation of the first kind, the bigger the 
degree of the ill-posedness of the equation \cite{HT}. Another example is the reconstruction of a penetrable obstacle from far field pattern of a scattered single incident wave.  The uniqueness has been proved in case of some obstacles with corners \cite{Els,Bla}, but in the general case it is open. 

As a third example, let us mention a problem to reconstruct a space-dependent source function  of a fractional diffusion equation from a measured state at a final time value $t=T$ over the space. The solution of this problem is unique \cite{JinRund}. If an initial state of the process is relatively less smooth than the source, then the final data contain sufficient information 
to recover simultaneously the source and an order of the fractional derivative $\alpha$ \cite{JKIP}. 
The proof of uniqueness is decomposed into 2 stages: 1) uniqueness of $\alpha$ and 2) uniqueness of the source. 
Similar results were obtained for inverse problems for fractional wave equations \cite{Liao} and problems with
unknown final time $T$ \cite{Kian1}.

In \cite{KJM} an inverse problem to reconstruct a time- and space dependent source $f$ of a  generalized fractional diffusion equation by means of overdetermination in a left neighborhood $(t_0,T)$ of the final time value $T$  was considered. It was shown that given values of $f$ and the
 state $u$   for  $t\in (t_0,T)$,  the history of  $f$  is uniquely recovered for $t\in (0,t_0)$. 
  Let us suppose that we have a possibility to 
choose $f$ for $t\in (t_0,T)$ and ask: can we do  it so that the measured  $u$ in $(t_0,T)$ contains sufficient information to recover simultaneously the history of $f$  and parameters of the equation?
In this paper we will  show that this is possible provided $f$ is a non-analytic with respect to $t$ function of special form, namely such that $f\equiv 0$ for $t\in (t_0,t_1)$ and $f\not\equiv 0$  for $t\in (t_1,T)$, where $t_1$ is some number between $t_0$ and $T$. A proof of uniqueness consists of 2 stages. Firstly, thanks to the assumed irregularity of $f$,  we can exclude the unknown history and prove the uniqueness of the parameters. 
After that we prove the uniqueness of the history of source. In addition, we will study the uniqueness of a problem to identify a kernel  contained in the generalized fractional diffusion equation from a measured functional of the state for $t\in (t_0,T)$ in case  the source is unknown for $t\in (0,t_0)$.

Let us give an overview of other studies in the field of inverse source problems for fractional diffusion equations. Several papers \cite{ad1,ad2,ad3,ad4} deal with reconstruction of time-dependent components of the sources 
making use of different measurements over time. The paper \cite{ad5} is concerned with the reconstruction of the source function that
depends on time and part of spatial variables from boundary measurements over the
time. In \cite{ad6,ad7,ad8} determination source functions in a form of separated variables under overdetermination in subdomains or portions of boundary is studied.

\section{Physical background. Formulation of problems and overview of results} \label{sec:back}

Fractional (and generalized fractional) diffusion equations are widely used to model anomalous diffusion processes in biology, social sciences, engineering sciences etc. \cite{Che,Fro,Sun}. There are several ways to derive such equations. 
Some methods are based 
on continuous time random walk \cite{Sch} or particle flows \cite{Fro}. A more classical approach \cite{Pov} presumes  a constitutive relation with memory
$
Q(t,x)=-{\partial\over \partial t}\int_0^t M(t-\tau) \nabla u(\tau,x)d\tau,
$
where $t>0$ is the time, $x\in \R^d$, $d\in\N$, is a space variable, $Q$ is the flux, $u$ is a state variable and $M$ is a memory kernel. Plugging this relation into the conservation equation 
${\partial \over\partial t}u+{\rm div}\, Q=f$, where $f$ is the source, we reach the following diffusion equation:
\beq\label{ph1}
{\partial \over\partial t} u(t,x)-{\partial\over \partial t}M*\Delta u(t,x)=f(t,x),
\eeq
where $*$ stands for the time convolution, i.e. $$v_1*v_2(t)=\int_0^t v_1(t-\tau)v_2(\tau)d\tau,$$ and $\Delta$ is the Laplacian. 

The operator ${\partial\over \partial t}M*$ is a generalized fractional derivative with the kernel $M$. 
In the case of the usual fractional diffusion, $M(t)=c {t^{\alpha-1}\over \Gamma(\alpha)}$, $0<\alpha<1$, $c>0$, Then ${\partial\over \partial t}M*=c D^{1-\alpha}$, where 
$D^{1-\alpha}$ is the Riemann-Liouville fractional derivative of the order $1-\alpha$
 and the equation \eqref{ph1} takes the form ${\partial \over\partial t} u-c D^{1-\alpha}\Delta u=f$. Such an equation is self-similar, i.e. rescaling the time does not change the type of memory. However, 
several anomalous diffusion processes are not self-similar. Then other memory functions are used. One example is the kernel of distributed fractional derivative
\beq\label{M1}
M(t)=\int_0^1 {t^{\alpha-1}\over\Gamma(\alpha)} dp(\alpha),
\eeq
where $p(\alpha)$ is a nonnegative Borel measure \cite{Mai,Sok}. This includes as particular cases the kernels of  multiterm fractional derivatives $M(t)=\sum_{l=1}^n \varkappa_l {t^{\alpha_l-1}\over\Gamma(\alpha_l)}$, $\varkappa_l>0$, 
$1>\alpha_1>\ldots>\alpha_n>0$, and the kernels of the form $M(t)=\int_0^1 \varkappa(\alpha) {t^{\alpha-1}\over\Gamma(\alpha)} d\alpha$, $\varkappa\ne 0$, $\varkappa\ge 0$. Another 
example is the kernel of tempered fractional derivative \cite{Gaj,Wu}
\beq\label{M2}
M(t)=c e^{-\lambda t}{t^{\alpha-1}\over \Gamma(\alpha)},\; 0<\alpha<1,\,  \lambda>0,\, c>0.
\eeq

The kernels \eqref{M1} and \eqref{M2} belong to the following space of integrable at $t=0$ completely monotonic functions: 
\beq\label{CM}
\begin{split}
&\mathcal{CM}=\big\{ M\, :\,  M\in L_{1,loc}(0,\infty)\cap C^\infty(0,\infty),\; 
\\
&\qquad \quad (-1)^n M^{(n)}(t)\ge 0,\; t>0,\; n=0,1,2,\ldots\big\}.
\end{split}
\eeq
Moreover, they satisfy the condition $\lim_{t\to 0^+}M(t)=\infty$. 

If $M\in \mathcal{CM}$ and $\lim_{t\to 0^+}M(t)=\infty$ then the 
Sonine equation 
\beq\label{son}
M*K(t)=1, \quad t>0,
\eeq
has the unique solution in $K\in L_{1,loc}(0,\infty)$ and this solution has the properties $K\in \mathcal{CM}$, $\lim_{t\to 0^+}K(t)=\infty$ (\cite{Gri1},  Theorem 3) 
In such  a case the equation \eqref{ph1} can be transformed 
to the following equation that contains the explicit Laplacian:
\beq\label{ph2}
 K*{\partial \over\partial t}u(t,x)-\Delta u(t,x)=K*f(t,x).
 \eeq
If $M(t)=c {t^{\alpha-1}\over \Gamma(\alpha)}$, $0<\alpha<1$,  then the solution of \eqref{son} is $K(t)={1\over c} {t^{-\alpha}\over \Gamma(1-\alpha)}$ and
\eqref{ph2} involves the Caputo fractional derivative of the order $\alpha$. To the author's opinion, the form \eqref{ph2} of the fractional generalized diffusion equation prevails in mathematical literature. 
If $M$ (and therefore also $K$) are known, then it is possible to switch from \eqref{ph1} to \eqref{ph2} and vice versa. But if $M$ is unknown then it is preferable to work 
with the equation \eqref{ph1}. 

If in addition to the diffusion a linear reaction occurs then the term $\Delta u$ in \eqref{ph1} is complemented with an addend $au(x)$, where $-a$ is the speed of the reaction \cite{Fro,KJMMA}. Moreover, in 
 models of anisotropic 
anomalous diffusion more general elliptic operators of the form\break $\sum_{l,m=1}^d a_{lm}{\partial^2\over\partial x_l\partial x_m}+\sum_{l=1}^d a_{l}{\partial \over\partial x_l}$ occur  \cite{Men}. 
In this paper we will consider the following equation: 
\beq\label{ph3}
{\partial \over\partial t} u(t,x)-{\partial\over \partial t}M*\mathscr{A}[\mathbf{a}] u(t,x)=f(t,x),
\eeq
where
\beq\label{Abasic} 
&&\hskip -9truemm \mathscr{A}[\mathbf{a}]v(x)\!=\!\sum_{l,m=1}^d\!\! a_{lm}(x){\partial^2\over\partial x_l\partial x_m}v(x)\!+\!\sum_{l=1}^d a_{l}(x){\partial \over\partial x_l}v(x)\!+\!a(x)v(x),\;
\\ \nonumber
&&\hskip -7truemm\mathbf{a}=\big((a_{lm})|_{l,m=1,\ldots,d},(a_l)|_{l=1,\ldots,d},a\big).
\eeq

In equations modelling nonlocal in space diffusion processes fractional or generalized fractional elliptic operators  are involved  \cite{Sok,Cap,J}. Another equation that we will consider in our paper is the following one:
\beq\label{ph4}
{\partial \over\partial t} u(t,x)-{\partial\over \partial t}M*A_{\varrho,a} u(t,x)=f(t,x),
\eeq
where $A_{\varrho,a}=-\int_0^1 (-\Delta -aI)^\beta d\varrho(\beta)$ is a distributed fractional elliptic operator defined in Subsection \ref{Ex}.

Let $u$ solve either \eqref{ph3} or \eqref{ph4} in $(0,T)\times \Omega$, where $T>0$, $\Omega\subset \R^d$. Assume that $f$ and $u$ are sufficiently smooth, $u(t,\cdot)$ belongs to a set 
where the space operator (i.e. $\mathscr{A}[\mathbf{a}]$ or $A_{\varrho,a}$, respectively) is injective and $M\in \mathcal{CM}_\dagger$, where
\beq
\label{CM+}
\begin{split}
&\mathcal{CM}_\dagger=\{M\in \mathcal{CM}\, :\, \mbox{the Laplace transform of $M$}
\\
&\quad\mbox{cannot be meromorphically extended to the whole complex plane}\}.
\end{split}
\eeq
Moreover, suppose that $f|_{(t_0,T)\times\Omega}$ and $u|_{(t_0,T)\times\Omega}$ are given, where $t_0$ is some number in $(0,T)$. Then  history of  $f$  and $u$ is uniquely recovered in $(0,t_0)\times\Omega$ \cite{KJM}. 

This result is valid in case of the kernels \eqref{M1} and \eqref{M2}, because they  belong to $\mathcal{CM}_\dagger$ \cite{KJM}.

In the present paper we will find conditions for $f|_{(t_0,T)\times\Omega}$ such that in addition to the history of $f$  and $u$  
parameters of the equation  (i.e. $M$ and elliptic operators) are uniquely recovered by
$f|_{(t_0,T)\times\Omega}$ and $u|_{(t_0,T)\times\Omega}$.

We will work with inverse problems for an abstract equation\break ${d\over dt}[u(t)-AM*u(t)]=f(t)$, $t\in (0,T)$,  in a Banach space $X$, where $A$ is a sectorial operator. 
In next section we will deduce some auxiliary results and  in Subsection \ref{Bas} we will prove Theorem \ref{th1} that establishes the uniqueness for a problem to 
determine simultaneously $f$, $u$ and $MA$ provided $f|_{(t_0,T)}$ and 
$u|_{(t_0,T)}$ are given and $f|_{(t_0,T)}$
satisfies certain restrictions. In Subsection \ref{Ex} we will apply this result to equations ${\partial \over\partial t} [u(t,x)-\mathscr{A}[\mathbf{a}]M* u(t,x)](t,x)=f(t,x)$ and
${\partial \over\partial t} [u(t,x)-A_{\varrho,a}M*u(t,x)]=f(t,x)$ that are weaker forms of \eqref{ph3} and \eqref{ph4}, respectively. There Corollaries \ref{coro1} - \ref{coro3} provide 
the uniqueness of parameters of 
$\mathscr{A}[\mathbf{a}]$ and $A_{\varrho,a}$, too.  
 In Subsection \ref{ss1} 
we will study uniqueness of a problem to determine simultaneously $f$, $u$   and $M$ from the same data in case of less restrictions on $f|_{(t_0,T)}$ (Theorem \ref{thM1}) and  in Subsection \ref{ss2} we will prove the uniqueness for a problem to recover $M$ provided $f$ and a functional of $u$ are given in $(t_0,T)$ (Theorem \ref{thmviim}). In the latter case the unique reconstruction of history of $f$ and $u$ may in general not be possible. The final Section \ref{conc} contains 
 conclusions.

\section{Preliminaries}\label{sec:prel}

Let $X$ be a complex Banach space, $X^*$ its dual and $\mathcal{B}(X)$ the space of linear bounded operators in $X$. 
We will denote the norms in $X$ and $\mathcal{B}(X)$ by $\|\cdot\|$   and  the pairing $X^*\times X\mapsto \C$  by $\langle\cdot,\cdot\rangle$. 

Further, let $A\, :\, D(A)\mapsto X$ be a linear closed densely defined operator in $X$. Let us denote by $X_A$ the set $D(A)$ endowed with the graph norm 
$\|{\rm x}\|_A=\|{\rm x}\|+\|A{\rm x}\|$. 
Since $A$ is closed, $X_A$ is a Banach space with respect to this norm.

Suppose that $M\in L_{1,loc}(0,\infty)$ and consider the following equation:
\beq\label{ab1}
v(t)-AM*v(t)=g(t),\;\; t\in [0,T]
\eeq
in the space $X$. 

A family of operators $S\, :\, [0,\infty)\mapsto {\mathcal B}(X)$ is called the resolvent of \eqref{ab1} if $S(t)$ is strongly continuous in $[0,\infty)$ and
\beq\nonumber
&&S(t){\rm x}\in D(A),\; AS(t){\rm x}=S(t)A{\rm x}, \; t\in [0,\infty),\;\;\mbox{for}\;\;{\rm x}\in D(A),
\\ \label{resolv}
&&S(t){\rm x}-M*AS(t){\rm x}={\rm x}, \; t\in [0,\infty),\;\;\mbox{for}\;\;{\rm x}\in D(A).
\eeq

We formulate a lemma regarding the equation  \eqref{ab1}.

\begin{lemma}\label{pruss}{\rm (\cite{Pruss}, Proposition 1.2)} Let \eqref{ab1} have a resolvent. Then the following assertions are valid.
\begin{description}
\item{\rm (i)} (uniqueness) If $v\in C([0,T];X)$ satisfy $M*v\in C([0,T];X_A)$ and solve $v(t)-AM*v(t)=0$, $t\in [0,T]$, then $v=0$. 
\item{\rm (ii)}  If $g\in W_1^1((0,T);X)$ then the function
\beq\label{pruss1}
v(t)=S(t)g(0)+S*g'(t)
\eeq
belongs to $ C([0,T];X)$, satisfies $M*v\in C([0,T];X_A)$ and is a solution of \eqref{ab1}.
\item{\rm (iii)}  If  $g\in W_1^1((0,T);X_A)$ then $v$ defined by \eqref{pruss1} belongs to $C([0,T];X_A)$ and is a solution of the stronger equation 
$$v(t)-M*Av(t)=g(t),\;\; t\in [0,T].$$
\end{description}
\end{lemma}

Next we are going to establish sufficient conditions for the existence of the resolvent of \eqref{ab1} and its analyticity. 
To this end we introduce additional notation. Denote
$$
\Sigma (\omega,\theta)=\{z\in\C\setminus\{\omega\}\, :\, |\arg(z-\omega)|<\theta\},\; \omega\in\R, \,\theta\in (0,\pi].
$$
Let   $\omega\in\R, \,\theta\in (0,\pi/2]$. The say that the operator $A$ belongs to a class  $\mathcal{S}(\omega,\theta)$ if
\beqst
 \rho(A)\supset 
\Sigma (\omega,{\pi\over 2}+\theta) ,\;\;
\exists C>0\, :\, \|(\lambda I-A)^{-1}\|\le {C\over |\lambda-\omega|},\, \lambda\in \Sigma (\omega,{\pi\over 2}+\theta).
\eeqst
We mention that $\mathcal{S}(\omega,\theta)$ is the class of sectorial operators that generate analytic semigroups \cite{Lun}.

Let $\psi \, :\, [0,\infty)\mapsto Y$, where $Y$ is a complex Banach space and $e^{-\sigma t}\psi\in L_1((0,\infty);X)$ for some $\sigma\in\R$. By $\widehat \psi$ we denote the Laplace transform of $\psi$, i.e. 
$$
\widehat \psi (s)=\int_0^\infty e^{-st} \psi(t)dt,\;\; s\in\C\, :\, {\rm Re}\, s>\sigma.
$$

\begin{lemma}\label{lemM}
Let $M\in \mathcal {CM}$. Then $\widehat M(s)$ is analytically extendable to $\Sigma (0,\pi)$. Moreover, for any $\theta\in (0,\pi)$  there exists a
 nonincreasing function $\mu_{M,\theta}\, :\, (0,\infty)\mapsto (0,\infty)$ such that 
$\lim\limits_{\rho\to\infty}\mu_{M,\theta}(\rho)= 0$ and $|\widehat M(s)|\le \mu_{M,\theta}(|s|)$, $|\arg s|\le \theta$.
\end{lemma}

\noindent{\it Proof}. Since $M$ is completely monotonic, due to the Bernstein's theorem, there exists a nonnegative Borel measure $q$ such that $M(t)=\int_0^\infty e^{-t\tau}dq(\tau)$. 
The inclusion $M\in L_{1,loc}(0,\infty)$ implies $\int_{\delta}^\infty {1\over\tau} dq(\tau)<\infty$ for any $\delta>0$.
We have 
\beq\label{lemM1}
\widehat M(s)= \int_0^\infty e^{-st}\int_0^\infty e^{-t\tau}dq(\tau)=\int_0^\infty{1\over s+\tau}dq(\tau),\quad {\rm Re}\, s>0.
\eeq
Let $\theta\in (0,\pi)$. For any  $s$ such that $|\arg s|\le \theta$ and $\tau>0$
we have $${\tau\over |s+\tau|}={\tau\over \sqrt{|s|^2+2|s|\tau \cos \arg s+\tau^2}}\le {\tau\over \sqrt{|s|^2+2|s|\tau \cos \theta+\tau^2}}\le c_\theta,$$ where 
$c_\theta=1$ if $\theta\le {\pi\over 2}$ and $c_\theta=\sqrt{1\over 1-\cos^2\theta}$ if $\theta>{\pi\over 2}$. Moreover, if 
$0<\tau<{|s|\over 2}$ then $|s+\tau|>{|s|\over 2}$. Therefore, we obtain  
\beqst
\int_0^\infty{1\over |s+\tau|}dq(\tau)\le {2\over |s|}\int_0^{|s|\over 2}  dq(\tau)+c_\theta\int_{|s|\over 2}^\infty {1\over\tau} dq(\tau)<\infty, 
\;\; |\arg s|\le \theta.
\eeqst
 Therefore, 
$\int_0^\infty{1\over s+\tau}dq(\tau)$  defines an analytic function of $s$ in $\Sigma (0,\theta)$. 
Since $\theta\in (0,\pi)$ is arbitrary, $\int_0^\infty{1\over s+\tau}dq(\tau)$ is analytic $\Sigma (0,\pi)$. This with \eqref{lemM1} proves that $\widehat M(s)$ is analytically extendable to  $\Sigma (0,\pi)$. Further, let
 us define $\mu_{M,\theta}^0(\varrho)=
\int_0^\infty P(\varrho,\tau)dq(\tau)$,  where
$$
P(\varrho,\tau)=\left\{\begin{array}{ll}{1\over \sqrt{\varrho^2+2\varrho\tau \cos \theta+\tau^2}}\;\;&\mbox{if}\;\; \tau>{\varrho\over 2}
\\[2ex]
{2\over \varrho}\;\;&\mbox{if}\;\; 0<\tau<{\varrho\over 2}.\end{array}\right.
$$
Then $|\widehat M(s)|\le \mu_{M,\theta}^0(|s|)$ if $|\arg s|\le \theta$. The function
 $P(\varrho,\tau)$ approaches $0$ as $\varrho\to\infty$ for any $\tau>0$ and is bounded by the integrable function $B(\tau)=\left\{\begin{array}{ll}\max\{c_\theta;1\}{1\over\tau}\;&\mbox{if}\;\tau>1\\
 1\;&\mbox{if}\;0<\tau<1\end{array}\right.$ for $\varrho>2$. Due to the 
 dominated convergence theorem, $\lim\limits_{\rho\to\infty}\mu_{M,\theta}^0(\rho)= 0$. Finally, we define $\mu_{M,\theta}(\rho)=\sup\limits_{\lambda>\rho}\mu_{M,\theta}^0(\lambda)$. 
 The proof is complete. \hfill $\Box$

Similarly to the resolvent of \eqref{ab1} we can define a resolvent of the perturbed equation 
$
v(t)-AM*v(t)-b*v(t)=g(t),
$
where $b\in L_{1,loc}(0,\infty)$.
This is a family of operators $S\, :\, [0,\infty)\mapsto {\mathcal B}(X)$ that is strongly continuous in $[0,\infty)$, 
$S(t){\rm x}\in D(A),\; AS(t){\rm x}=S(t)A{\rm x}, \; t\in [0,\infty)\;\;\mbox{for}\;\;{\rm x}\in D(A)$ and 
$
S(t){\rm x}-M*AS(t){\rm x}-b*S(t){\rm x}={\rm x}, \, t\in [0,\infty)\;\;\mbox{for}\;\;{\rm x}\in D(A).
$

\begin{lemma}\label{lempr}
Let $A_0\in \mathcal{S}(0,\theta)$, $\theta\in (0,{\pi\over 2}]$, $M\in \mathcal{CM}$, $b\in L_{1,loc}(0,\infty)$. 
Then the equation 
\beq\label{abv0}
v(t)-A_0M*v(t)-b*v(t)=g(t)
\eeq
  has a resolvent $S(t)$. It can be represented
by the series
\beq\label{abp1}
S(t)=\sum_{n=0}^\infty G_n(t),
\eeq
where $G_n\, :\, [0,\infty)\mapsto {\mathcal B}(X)$, $k\in\{0\}\cup \N$, is a family of operators whose Laplace transforms can be analytically extended to the set $\Sigma(0,{\pi\over 2}+\theta)$
and have the formulas
\beq \nonumber
&&\hskip -1truecm \widehat G_n(s)=(1+\widehat r(s))(\widehat r(s))^n{1\over s}{(I-\widehat M(s)A_0)^{-1}}
\\ \label{abp2}
&&\times ( \widehat M(s)A_0(I-\widehat M(s)A_0)^{-1})^n,\quad s\in \Sigma(0,{\pi\over 2}+\theta),
\eeq
and $r$ is the solution of the integral equation $r=b+b*r$. Moreover,   
\beq\label{abp3}
\|(I-\widehat M(s)A_0)^{-1} \|\le {C_2}, \; s\in \Sigma(0,{\pi\over 2}+\theta) 
\eeq
with some constant $C_2$.
\end{lemma}

\noindent{\it Proof}. The assumption $A_0\in \mathcal{S}(0,\theta)$, $\theta\in (0,{\pi\over 2}]$, implies that $A_0$ generates a semigroup that is analytic and bounded in 
$\Sigma(0,\theta)$ \cite{Lun}.  Due to this property and the assumption $M\in \mathcal{CM}$, Corollary 2.4 of \cite{Pruss} implies that \eqref{abv0} with $b=0$ has a resolvent  $S_0(t)$ that is bounded and analytic in 
$\Sigma(0,\theta)$. This fact with  Theorem 2.1 of \cite{Pruss} implies the relations ${1\over \widehat M(s)}\in \rho(A_0)$, $s\in \Sigma(0,{\pi\over 2}+\theta)$,  and the estimate \eqref{abp3}. 
Further, the existence of the resolvent $S(t)$ of  \eqref{abv0}  follows from the mentioned properties of $S_0(t)$ and  $b\in L_{1,loc}(0,\infty)$ from 
Theorem 2.3 of \cite{Pruss}.  The formulas \eqref{abp1} and \eqref{abp2} are contained in 
a proof of  the latter theorem.    \hfill $\Box$

\begin{lemma}\label{lem1} Let  $A\in \mathcal{S}(\omega,\theta)$ for some 
$\omega\in\R$, $\theta\in (0,{\pi\over 2}]$ and $M\in \mathcal{CM}$. 
Then \eqref{ab1} has a resolvent $S(t)$ that can be analytically extended to the sector $\Sigma(0,\theta)$. 
\end{lemma}

\noindent{\it Proof}. 
Let us  define 
 $A_0=A-\omega I$ and $b=\omega M$. Then \eqref{ab1} takes the form of \eqref{abv0}, so the existence of the resolvent $S(t)$ of \eqref{ab1} follows from Lemma \ref{lempr}. It remains to show that $S(t)$
 is analytically extendable  to  $\Sigma(0,\theta)$.

Let us estimate $\widehat G_n(s)$ given by \eqref{abp2}. For the term $M(s)A_0(I-\widehat M(s)A_0)^{-1}$ due to \eqref{abp3} we have 
\beq\nonumber
&&\| \widehat M(s)A_0(I-\widehat M(s)A_0)^{-1}\|=\|(I-\widehat M(s)A_0)^{-1}-I\| 
\\ \label{abp4}
&&\quad \le C_2+1,\quad s\in \Sigma(0,{\pi\over 2}+\theta).
\eeq
The Laplace transform $\widehat r(s)$ of the function  $r$ that solves the equation $r=b+b*r$ with $b=\omega M$ has the formula 
$$
\widehat r(s)={\omega \widehat M(s)\over 1-\omega\widehat M(s)}.
$$
Let $\theta'\in (0,\theta)$. Then $\cos \theta'>0$ and 
for any  $\eta>0$ and  $s\in \Sigma(\eta,{\pi\over 2}+\theta')$ we have $|s|\ge \eta \cos\theta'$. Therefore, 
in view of Lemma \ref{lemM}, it holds $|\omega\widehat M(s)|\le |\omega|\mu_{M,\theta'}(\eta \cos\theta')$ for $s\in \Sigma(\eta,{\pi\over 2}+\theta')$, $\eta>0$. 
Since $\mu_{M,\theta'}(\eta \cos\theta')\to 0$  as $\eta\to\infty$, there exists a sufficiently big $\eta=:\omega_2$  such  that 
\beq\label{abp5}
|\omega\widehat M(s)|<1\quad\mbox{and}\quad    |\widehat r(s)|\le {1\over 2(C_2+1)}\quad \mbox{for}\; s\in \Sigma(\omega_2,{\pi\over 2}+\theta').
\eeq
Using \eqref{abp3} - \eqref{abp5} in \eqref{abp2} we obtain $\|\widehat G_n(s)\|\le {C_3\over 2^n |s|},\quad s\in  \Sigma(\omega_2,{\pi\over 2}+\theta')$ 
with some constant $C_3$  and since ${1\over |s|}\le ({1\over \cos\theta'}+1){1\over |s-\omega_2|}$, $s\in  \Sigma(\omega_2,{\pi\over 2}+\theta')$, we have
\beq\label{abp6} 
\|\widehat G_n(s)\|\le {C_4\over 2^n |s-\omega_2|},\quad s\in  \Sigma(\omega_2,{\pi\over 2}+\theta'),
\eeq
with a constant $C_4$. 
 Let  us define an extension of $G_n(t)$  to complex arguments $z$ by the following integral:
\beq\label{abp7}
G_n(z)={1\over 2\pi i}\int_{\gamma_R}e^{zs}\widehat G_n(s)ds,
\eeq
where ${\rm Re}z>0$, $R={1\over |z|}$, $\gamma_R=\gamma_R^1\cup\gamma_R^2\cup\gamma_R^3$, $\gamma_R^1=\{z=\omega_2+\rho e^{-i({\pi\over 2}+\theta)},\; \infty>\rho>R\}$, 
$\gamma_R^2=\{z=\omega_2+R e^{i\tau},\; -{\pi\over 2}-\theta<\tau<{\pi\over 2}+\theta\}$, $\gamma_R^3=\{z=\omega_2+\rho e^{i({\pi\over 2}+\theta)},\; R<\rho<\infty\}$.
By means of a standard  procedure (see e.g. \cite{Pruss}, p. 52) and \eqref{abp6} we deduce that 
this integral converges absolutely for $z\in \Sigma(0,\theta')$, and satisfies the estimate
\beq\label{abp8}
\|G_n(z)\|\le  {C_{5,\theta_1} C_4\over 2^n}e^{\omega_2 {\rm Re}z},\quad z\in \Sigma(0,\theta_1),
\eeq
 where $\theta_1$ is any number in between $0$ and $\theta'$  and $C_{5,\theta_1}$ is a constant depending on $\theta_1$. 
 Therefore, $G_n(z)$ is analytic in $\Sigma(0,\theta')$. Finally, we define the extension of $S(t)$ to $\Sigma(0,\theta')$ by 
$$
S(z)=\sum_{n=0}^\infty G_n(z).
$$
Due to \eqref{abp8}, this series converges uniformly in compact subsets of $\Sigma(0,\theta_1)$. Therefore, $S(z)$ is analytic in $\Sigma(0,\theta_1)$. Since $\theta_1\in (0,\theta')$ 
is arbitrary and $\theta'\in (0,\theta)$ arbitrary,  $S(z)$ is analytic in $\Sigma(0,\theta)$. \hfill $\Box$

\begin{lemma}\label{lemanal}
Let  $\theta\in (0,{\pi\over 2}]$ and  $S(z)\, :\, \Sigma(0,\theta)\mapsto {\mathcal B}(X)$ be analytic. Moreover, let 
 $f\in L_{1,loc}( (0,\infty);X)$ and $t_0>0$. Then $$h(z)=\int_0^{t_0}S(z-\tau)f(\tau)d\tau$$ is analytic in $\Sigma(t_0,\theta)$. 
\end{lemma}

\noindent{\it Proof}. Let $t_1>t_0$, $z,z_1\in \Sigma(t_1,\theta)$, $\phi\in X^*$.  We have
\beqst
&&{\langle\phi,h(z_1)\rangle-\langle\phi,h(z)\rangle\over z_1-z}=\int_0^{t_0}H(z,z_1,\tau)d\tau,
\\
&&H(z,z_1,\tau)=\left\langle\phi, {1\over z_1-z} (S(z_1-\tau)-S(z-\tau))f(\tau)\right\rangle.
\eeqst
For a.e. $\tau\in (0,t_0)$ we have $\lim\limits_{z_1\to z}H(z,z_1,\tau)=\left\langle\phi, S'(z-\tau)f(\tau)\right\rangle$. On the other hand, $H(z,z_1,\tau)$ can be represented in the form
$H(z,z_1,\tau)=\left\langle\phi, {1\over z_1-z}\int_{\gamma_{z,z_1}}S'(y-\tau)f(\tau)dy\right\rangle$, 
where $\gamma_{z,z_1}$ is the straight path from $z$ to $z_1$. Thus, in case $|z_1-z|\le \varepsilon$, $\tau\in (0,t_0)$, 
\beqst
|H(z,z_1,\tau)|\le \|\phi\| \max_{y\in \overline{\Sigma(t_1-t_0,\theta)}\cap \{y\, :\, {\rm Re}\,y\le {\rm Re}\, z+\varepsilon\}}\|S'(y)\| \|f(\tau)\|.
\eeqst
This shows that $H(z,z_1,\tau)$ is bounded by an integrable in $(0,t_0)$ function of $\tau$ for $z, z_1$ such that $ |z_1-z|\le \varepsilon$. By dominated convergence theorem, 
${\langle\phi,h(z_1)\rangle-\langle\phi,h(z)\rangle\over z_1-z}\to \int_0^{t_0}\left\langle\phi, S'(z-\tau)f(\tau)\right\rangle d\tau$ as $z_1\to z$. This shows that 
$\langle\phi,h(z)\rangle$ is differentiable in $\Sigma (t_1,\theta)$, hence analytic in $\Sigma (t_1,\theta)$. Since $\phi\in X^*$ and $t_1>t_0$ are arbitrary, 
$h(z)$ is analytic in $\Sigma(t_0,\theta)$. \hfill $\Box$

\medskip
Finally, we prove a lemma concerning an abstract weak analogue of the equations \eqref{ph3} and \eqref{ph4}. 

\begin{lemma}\label{dplem}
Let $A\in \mathcal{S}(\omega,\theta)$ for some 
$\omega\in\R$, $\theta\in (0,{\pi\over 2}]$, $M\in \mathcal{CM}$, $f\in L_1((0,T);X)$ and $u_0\in X$. Then there exists a unique function $u\in C([0,T];X)$ that satisfies 
$M*u\in C([0,T];X_A)$,   
 $u-AM*u\in W_1^1((0,T);X)$, has the initial value $u(0)=u_0$ and solves the equation 
 \beq\label{dpeq}
{d\over dt}\left[u(t)-AM*u(t)\right]=f(t),\quad a.e.\;  t\in (0,T).
\eeq
\end{lemma}

\noindent{\it Proof}. Let us consider the equation
\beq\label{dplem1}
v(t)-AM*v(t)=1*f(t)+v_0,\;\; t\in [0,T].
\eeq
 By Lemma \ref{lem1}, this equation has the resolvent $S$ and by Lemma \ref{pruss} (ii), the function 
\beq\label{dplem2}
v(t)=S(t)u_0+S*f(t)
\eeq
belongs to $C([0,T];X)$, satisfies $M*v\in C([0,T];X_A)$ and solves \eqref{dplem1}. Since the right-hand side of \eqref{dplem1} belongs to $W_1^1((0,T);X)$, we have 
$v-AM*v\in W_1^1((0,T);X)$. Moreover, ${d\over dt}\left[v(t)-AM*v(t)\right]=f(t)$, a.e. $t\in (0,T)$ and from \eqref{dplem2} we deduce $v(0)=u_0$. 
Therefore, $u=v$ is a function that meets the requirements of the lemma. This proves the existence assertion. To prove the uniqueness, suppose that $u\in C([0,T];X)$ satisfies 
$M*u\in C([0,T];X_A)$,   
 $u-AM*u\in W_1^1((0,T);X)$,  $u(0)=0$ and solves ${d\over dt}\left[u(t)-AM*u(t)\right]=0$, a.e. $t\in (0,T)$. The latter relation implies $u(t)-AM*u(t)=c$, $t\in [0,T]$, where 
 $c\in X$. Due to Lemma \ref{pruss} (ii), $u(t)=S(t)c$. Using the condition $u(0)=0$ we get $c=0$. Lemma \ref{pruss} (i) implies $u=0$. This proves the uniqueness assertion.
 \hfill $\Box$

\section{Inverse problem to determine the operator, kernel, source and state}\label{sec:opker}

\subsection{Abstract inverse problem}\label{Bas}

Let $u$ solve \eqref{dpeq}. 
We pose the following inverse problem: given $u(t)$, $f(t)$, $t\in(t_0,T)$, where $t_0$ is some number between $0$ and $T$,  determine $MA$ and $f(t)$, $u(t)$ in the whole interval $(0,T)$. 
In  the sequel we prove the uniqueness for this problem.

\begin{theorem}\label{th1} 
Let $M_1\in \mathcal {CM}_\dagger$,  $M_2\in \mathcal {CM}$, $A_j\, :\, D(A_j)\mapsto X$, $j=1,2$ - 
 linear closed densely defined operators in a complex Banach space $X$,  $A_j\in \mathcal{S}(\omega_j,\theta_j)$
 for some 
$\omega_j\in\R$, $\theta_j\in (0,{\pi\over 2}]$ and $A_1$ - injective. Let $f_j$, $u_j$, $j=1,2$, satisfy
\beq\label{th1a1}
&&\hskip -5truemm f_j\in L_1((0,T);X),\;u_j\in C([0,T];X),\; 
\\ \label{th1a2}
&&\hskip -5truemm M_j*u_j\in C([0,T];X_{A_j}),\; u_j-A_jM_j*u_j\in W_1^1((0,T);X),
\\ \label{th1a3}
&&\hskip -5truemm {d\over dt}\left[u_j(t)-A_jM_j*u_j(t)\right]=f_j(t),\; a.e.\; t\in (0,T),
\\[1ex] \label{th1a4}
&&\hskip -5truemm f_1(t)=f_2(t)=:f(t)\; a.e.\; t\in (t_0,T)\; \mbox{for some $t_0\in (0,T)$},\; 
\eeq
where $f$ fulfills the condition
\beq
 \label{th1a5}
&&\hskip -5truemm f(t)=0\; a.e.\; t\in (t_0,t_1)\; \mbox{for some $t_1\in (t_0,T)$.}
\eeq
Moreover, assume that either 
\beq\label{th1v1}\left.
\begin{split}
&\mbox{$D(A_1)\cap D(A_2)$ is not empty and dense in $X_{A_j}$, $j=1,2$,}
\\
&f(t)=\sum_{i=1}^\infty \psi_i(t){\rm x}_i,\; t\in (t_1,T),\; \psi_i\in L_1(t_1,T),\; \\
&{\rm ess\, supp}\,\psi_i\subset [t_{i},t_{i+1}),\; t_1<t_2<\ldots <T,\;  
\\
&t_{i}=\min{\rm ess\, supp}\,\psi_i,\; {\rm span}\{{\rm x}_i\, :\, i\in\N\}  - \mbox{dense in  $X$}
\end{split}\right\}
\eeq
or
\beq\label{th1v2}\left.
\begin{split}
&X \; \mbox{is a Hilbert space},\;\; A_j  \; \mbox{is selfadjoint},\,\;\; 
\\
&\sigma (A_j)\; \mbox{consists of a countable number  of eigenvalues $\lambda_{jk}$, $k\in\N$}, \\
&\mbox{the eigenvectors ${\rm x}_k$, $k\in\N$, of $A_1$ and $A_2$ coincide, i.e. }
\\
&\mbox{$A_j{\rm x}_k=\lambda_{jk}{\rm x}_k$, $k\in \N$},\\
&\mbox{the system $({\rm x}_k)_{k\in\N}$ is orthonormal and complete in $X$},\;\;\\\
&\langle f(t),{\rm x}_k\rangle\not\equiv 0\;\mbox{in}\; (t_1,T),\:\;k\in\N.
\end{split}\right\}
\eeq
Finally, let 
\beq\label{th1a6}
u_1(t)=u_2(t),\; t\in (t_0,T).
\eeq
 Then $D(A_1)=D(A_2)$, there exists $c>0$ such that $cM_1=M_2$, $A_1=cA_2$, $f_1=f_2$ and $u_1=u_2$. 
\end{theorem}

In \eqref{th1v2},  $\langle\cdot,\cdot\rangle$ is interpreted as a scalar product in $X$.
We split most of the proof of this theorem in lemmas. 

\begin{lemma}\label{prop1} 
Let $M_j\in \mathcal {CM}$, $j=1,2$, $A_j\, :\, D(A_j)\mapsto X$, $j=1,2$ -  linear closed densely defined operators in a complex Banach space $X$ and $A_j\in \mathcal{S}(\omega_j,\theta_j)$
 for some 
$\omega_j\in\R$, $\theta_j\in (0,{\pi\over 2}]$. Moreover, let $f_j$, $u_j$, $j=1,2$, satisfy {\rm \eqref{th1a1} - \eqref{th1a4}, \eqref{th1a6}} and $f$ involved in  \eqref{th1a4} 
fulfill the condition \eqref{th1a5}.
 Then 
\beq\label{th1p4e}
\int_{t_1}^tS_1(t-\tau)f(\tau)d\tau=\int_{t_1}^tS_2(t-\tau)f(\tau)d\tau,\quad t\in [t_1,T],
\eeq
where $S_j$ is the resolvent  of the equation \eqref{ab1} with $M=M_j$ and $A=A_j$.
\end{lemma}

\noindent{\it Proof}. We have $u_j(t)-A_jM_j*u_j(t)=1*f_j(t)+u_j(0)$, $t\in [0,T]$. Due to Lemma \ref{pruss} we obtain
\beq\label{th1p1}
u_j(t)=S_j(t)u_j(0)+S_j*f_j(t),\quad t\in [0,T],\;\; j=1,2.
\eeq
Let $t\in [t_0,T]$. Subtracting the equation \eqref{th1p1} with $j=2$ from the one with $j=1$  and taking  \eqref{th1a4} and \eqref{th1a6} into account, we deduce
\beq\label{th1p2}
0=Y(t)+Z(t),\quad t\in [t_0,T],
\eeq
where
\beqst
&&\hskip-7truemm Y(t)=S_1(t)u_1(0)-S_2(t)u_2(0)
\\
&&\qquad +\int_0^{t_0}\left[S_1(t-\tau)f_1(\tau)d\tau-S_2(t-\tau)f_2(\tau)\right]d\tau,
\\
&&\hskip-7truemm Z(t)=\int_{t_0}^t\left[S_1(t-\tau)-S_2(t-\tau)\right]f(\tau)d\tau.
\eeqst
Since $f(t)=0$ a.e. $t\in (t_0,t_1)$,  we have $Z(t)=0$, $t\in [t_0,t_1]$. Using this in \eqref{th1p2} we get $Y(t)=0$,  $t\in [t_0,t_1]$. In view of  Lemmas \ref{lem1} and \ref{lemanal}, $Y(t)$ is analytic for $t>t_0$. By analytic continuation 
we obtain $Y(t)=0$,  $t\ge t_0$. Therefore, \eqref{th1p2} reduces to $Z(t)=0$, $t\in [t_0,T]$. This by $f(t)=0$ a.e. $t\in (t_0,t_1)$, implies \eqref{th1p4e}.
\hfill $\Box$

\medskip Lemma \ref{prop1} provides an independent equation \eqref{th1p4e}  that involves  $M_j$ and $A_j$ but not $u_j|_{(0,t_0)}$ and $f_j|_{(0,t_0)}$. 
Under certain assumptions, this equation implies $M_1A_1=M_2A_2$. We will show it in the next lemma.

\begin{lemma}\label{thm1l1}
Let the assumptions of Lemma {\rm \ref{prop1}} be satisfied for $M_j$, $A_j$, $j=1,2$. Let $f\in L_1((t_1,T);X)$ and \eqref{th1p4e} be valid where 
$S_j$ is the resolvent  of the equation \eqref{ab1} with $M=M_j$ and $A=A_j$. Moreover, assume that either \eqref{th1v1} or \eqref{th1v2} holds. 
Then $D(A_1)=D(A_2)$ and there exists $c>0$ such that $cM_1=M_2$ and $A_1=cA_2$.
\end{lemma}

\noindent{\it Proof}. 
Firstly, we consider the  case \eqref{th1v1}. Then from \eqref{th1p4e} we have
\beq\label{th1p4}
\int_{t_1}^t\left[S_1(t-\tau)-S_2(t-\tau)\right]\sum_{i=1}^\infty \psi_i(\tau){\rm x}_id\tau=0,\quad t\in [t_1,T].
\eeq
Let us show by induction that 
\beq\label{th1p6}
S_1(t){\rm x}_m=S_2(t){\rm x}_m,\quad t\ge 0,\;\; m=1,2,\ldots.
\eeq
To this end, we suppose that
 either $m_0=0$ or $m_0\in\N$ and \eqref{th1p6} holds for $m=1,\ldots,m_0$. The aim is to show that \eqref{th1p6} is valid for $m=m_0+1$, too.
Due to such a supposition, from \eqref{th1p4} in view of properties of $\psi_i$  we have 
\beqst\label{th1p4a}
\int_{t_{m_0+1}}^t\!\!\left[S_1(t-\tau)-S_2(t-\tau)\right]\sum_{i=m_0+1}^\infty \psi_i(\tau){\rm x}_id\tau=0,\;\; t\in [t_{m_0+1},T].
\eeqst
For $t\in [t_{m_0+1},t_{m_0+2}]$ this yields $\int_{t_{m_0+1}}^t\psi_{m_0+1}(\tau)[S_1(t-\tau)\break-S_2(t-\tau)]{\rm x}_{m_0+1}d\tau=0$. Let $\phi\in X^*$. Then 
\beqst\label{th1p5}
&&\int_{t_{m_0+1}}^t\!\!\psi_{m_0+1}(\tau)\langle\phi,\left[S_1(t-\tau)-S_2(t-\tau)\right]{\rm x}_{m_0+1}\rangle d\tau=0,
\\
&&\quad\quad\quad t\in [t_{m_0+1},t_{m_0+2}].
\eeqst
By \eqref{th1v1}, $\psi_{m_0+1}(t)$ does not vanish identically in an arbitrarily small right neighborhood of $t_{m_0+1}$. 
Applying Titchmarsh convolution theorem, we get $$\langle\phi,\left[S_1(t)-S_2(t)\right]{\rm x}_{m_0+1}\rangle=0,\;\; 
t\in [0,t_{m_0+2}-t_{m_0+1}].$$ Since $\phi\in X^*$ is arbitrary and the embedding $X\subseteq X^{**}$ in injective, we have $S_1(t){\rm x}_{m_0+1}=S_2(t){\rm x}_{m_0+1}$, $t\in [0,t_{m_0+2}-t_{m_0+1}]$, and by analytic continuation,
\eqref{th1p6} for $m=m_0+1$ follows. 

Since ${\rm span}\{{\rm x}_i\, :\, i\in\N\}$  is dense in  $X$, \eqref{th1p6} implies $S_1(t)=S_2(t)=:S(t)$, $t\ge 0$. Due to \eqref{resolv} we obtain
\beqst
S(t){\rm x}={\rm x}+M_j(t)*A_jS(t){\rm x},\quad t\ge 0,\; {\rm x}\in D(A_j),\; j=1,2.
\eeqst
This yields 
\beqst
v_{\rm x}(t):=S*[M_1(t)A_1-M_2(t)A_2]{\rm x}=0,\quad t\ge 0,\; {\rm x}\in D(A_1)\cap D(A_2).
\eeqst
From the left equality and Lemma \ref{pruss} we deduce that $v_{\rm x}$ is the solution of the following equation:
\beqst
v_{\rm x}(t)-A_jM_j*v_{\rm x}(t)=1*[M_1(t)A_1-M_2(t)A_2]{\rm x}, \quad t\ge 0,
\eeqst
for both $j=1,2$.
Therefore, $v_{\rm x}(t)=0$ implies $1*[M_1(t)A_1-M_2(t)A_2]{\rm x}=0$. This yields
\beq\label{th1p7}
M_1(t)A_1{\rm x}=M_2(t)A_2{\rm x},\quad t>0,\; {\rm x}\in D(A_1)\cap D(A_2).
\eeq
Let us define $c={M_2(1)\over M_1(1)}$. Then  setting $t=1$ in \eqref{th1p7}  we deduce
\beq\label{th1p77}
A_1{\rm x}=cA_2{\rm x},\; {\rm x}\in D(A_1)\cap D(A_2).
\eeq
Using \eqref{th1p77} in  \eqref{th1p7} we prove
 $cM_1=M_2$. Further, let ${\rm x}$ be an arbitrary element of $D(A_2)$. Since $D(A_1)\cap D(A_2)$ is dense in $X_{A_2}$, there exist ${\rm x}_n\in D(A_1)\cap D(A_2)$, $n\in\N$, 
such that ${\rm x}_n\to {\rm x}$ and $A_2{\rm x}_n\to A_2{\rm x}$ as $n\to\infty$. By \eqref{th1p77}, $A_1{\rm x}_n \to cA_2{\rm x}$ as $n\to\infty$. 
Since $A_1$ is closed, we have ${\rm x}\in D(A_1)$ and $A_1{\rm x}=cA_2{\rm x}$. This shows that 
$D(A_1)\supseteq D(A_2)$ and $A_1{\rm x}=cA_2{\rm x}$, ${\rm x}\in D(A_2)$. Similarly we show that $D(A_2)\supseteq D(A_1)$ and $A_2{\rm x}={1\over c}A_1{\rm x}$, ${\rm x}\in D(A_1)$. Consequently, 
$D(A_1)=D(A_2)$ and $A_1=cA_2$. 

Secondly, we consider the case \eqref{th1v2}. From \eqref{th1p4e} we have
\beq\label{th1p20eel}
\sum_{k=1}^\infty \int_{t_1}^t \!f_k(\tau)\langle\left[S_1(t\!-\!\tau)-S_2(t\!-\!\tau)\right]{\rm x}_k,{\rm x}_l\rangle d\tau=0,\; t\in [t_1,T],\;l\in\N,
\eeq
where $f_k(t)=\langle f(t),{\rm x}_k\rangle$.  The function $S_j(t){\rm x}_k$ solves the equation 
\beqst
S_j(t){\rm x}_k-M_j*A_jS_j(t){\rm x}_k={\rm x}_k, \quad t\ge 0.
\eeqst
Since $A_j$ commutes with $S_j$ and $A_j{\rm x}_k=\lambda_{jk}{\rm x}_k$, this implies
\beqst
S_j(t){\rm x}_k-\lambda_{jk}M_j*S_j(t){\rm x}_k={\rm x}_k, \quad t\ge 0.
\eeqst
Thus, 
\beq\label{th1p20}
\langle S_j(t){\rm x}_k,{\rm x}_l\rangle-\lambda_{jk}M_j*\langle S_j(t){\rm x}_k,{\rm x}_l\rangle=\delta_{kl}, \quad t\ge 0,
\eeq
where $\delta_{kl}$ is the Kronecker delta. The relation \eqref{th1p20} is a scalar Volterra equation of the second kind. In case $k\ne l$ it is homogeneous. Therefore,  
$\langle S_j(t){\rm x}_k,{\rm x}_l\rangle=0$, $t\ge 0$, $k\ne l$, and from \eqref{th1p20eel} we deduce
\beq\label{th1p21}
 \int_{t_1}^t f_k(\tau)\langle\left[S_1(t-\tau)-S_2(t-\tau)\right]{\rm x}_k,{\rm x}_k\rangle d\tau=0,\;\; t\in [t_1,T],\;k\in\N.
\eeq
Let us choose some $k\in\N$. By \eqref{th1v2}, $f_k(t)$ does not vanish in $(t_1,T)$. Denote $\tau_k=\min{\rm ess\, supp}\, f_k$. Then  \eqref{th1p21} reduces 
to\break  $\int_{\tau_k}^t f_k(\tau)\langle\left[S_1(t-\tau)-S_2(t-\tau)\right]{\rm x}_k,{\rm x}_k\rangle d\tau=0$,
$t\in [\tau_k,T]$. Moreover, $f_k(t)$  is not identically zero in an arbitrarily small right neighborhood of $\tau_k$. Applying Titchmarsh convolution theorem, we obtain
$\langle \left[S_1(t)-S_2(t)\right]{\rm x}_k,{\rm x}_k\rangle=0$, 
$t\in [0,T-\tau_k]$. By analytic continuation,
\beqst
\langle \left[S_1(t)-S_2(t)\right]{\rm x}_k,{\rm x}_k\rangle=0,\quad t\ge 0.
\eeqst
Using this relation and \eqref{th1p20} in case $l=k$ we deduce 
\beqst
(\lambda_{1k}M_1-\lambda_{2k}M_2)*\langle S_1(t){\rm x}_k,{\rm x}_k\rangle=0,\quad t\ge 0.
\eeqst
Since $S_1(0){\rm x}_k={\rm x}_k$ and  $S_1$ is strongly continuous, $\langle S_1(t){\rm x}_k,{\rm x}_k\rangle$ differs from zero in a right neighborhood of $0$. Using the Titchmarsh convolution theorem again,
we obtain 
\beqst
\lambda_{1k}M_1(t)=\lambda_{2k}M_2(t),\quad t>0.
\eeqst
We have chosen an arbitrary $k\in\N$. Therefore, this relation is valid for any $k\in\N$. Define $c={M_2(1)\over M_1(1)}$. Arguing as in the previous case we deduce
 $cM_1=M_2$ and $\lambda_{1k}=c\lambda_{2k}$, $k\in\N$. Let us define the set 
\beqst
D=\left\{ {\rm x}\in X\, :\, \sum_{k=1}^\infty |\lambda_{2k}|^2 |\langle {\rm x},{\rm x}_k\rangle|^2<\infty\right\}.
\eeqst
For any ${\rm x}\in D(A_1)$ we have $\langle A_1{\rm x},{\rm x}_k\rangle=\langle {\rm x},A_1{\rm x}_k\rangle=\lambda_{1k}\langle {\rm x},{\rm x}_k\rangle$. 
Therefore, $\|A_1{\rm x}\|^2=\sum_{k=1}^\infty |\lambda_{1k}|^2|\langle {\rm x},{\rm x}_k\rangle|^2=
c^2\sum_{k=1}^\infty |\lambda_{2k}|^2|\langle {\rm x},{\rm x}_k\rangle|^2<\infty$. Thus, ${\rm x}\in D$ and we have $D(A_1)\subseteq D$. Next let us choose some ${\rm x}\in D$. Since 
$\sum_{k=1}^\infty |\lambda_{2k}|^2 |\langle {\rm x},{\rm x}_k\rangle|^2<\infty$, the element  ${\rm y}:=\sum_{k=1}^\infty 
\lambda_{2k}\langle {\rm x},{\rm x}_k\rangle x_k$ is well-defined in $X$. 
Since ${\rm x}_k\in D(A_1)$, $k\in\N$, it holds 
$\hat {\rm x}_n:=\sum_{k=1}^n \langle {\rm x},{\rm x}_k\rangle x_k\in D(A_1)$. We deduce $\hat {\rm x}_n\to {\rm x}$ and $A_1\hat {\rm x}_n=\break \sum_{k=1}^n
 \lambda_{1k}\langle {\rm x},{\rm x}_k\rangle {\rm x}_k=c\sum_{k=1}^n
 \lambda_{2k}\langle {\rm x},{\rm x}_k\rangle {\rm x}_k\to c{\rm y}$ as $n\to\infty$.
Since $A_1$ is closed, we have ${\rm x}\in D(A_1)$. This shows that $D\subseteq D(A_1)$. Consequently, $D(A_1)=D$. Similarly we prove $D(A_2)=D$. Finally, 
the relation $A_1=cA_2$ follows from the equality $A_1{\rm x}=\sum_{k=1}^\infty \lambda_{1k}\langle {\rm x},{\rm x}_k\rangle {\rm x}_k=c\sum_{k=1}^\infty \lambda_{2k}\langle {\rm x},{\rm x}_k\rangle {\rm x}_k=cA_2{\rm x}$. \hfill $\Box$

\medskip In order to complete the proof of Theorem \ref{th1},  we will need one more lemma.

\begin{lemma}\label{thm1l2}
Let $M\in \mathcal {CM}_\dagger$, $A\, :\, D(A)\mapsto X$ -  an injective operator in a complex Banach space $X$ and $\varphi$, $u$ satisfy
   $ \varphi\in L_1((0,T);X)$, $u\in C([0,T];X)$, $M*u(t) \in D(A)$,  a.e. $t\in (0,T)$,
 $u-AM*u\in W_1^1((0,T);X)$ and solve the equation 
  \beq\label{dpeq1}
{d\over dt}\left[u(t)-AM*u(t)\right]=\varphi(t),\quad a.e.\;  t\in (0,T).
\eeq
Moreover, assume that for some $t_0\in (0,T)$ it holds $\varphi(t)=0$ a.e. $t\in (t_0,T)$ and $u(t)=0$, $t\in (t_0,T)$. Then $\varphi=0$ and $u=0$.
\end{lemma}

If $u$ is a sufficiently good function then from \eqref{dpeq1} we obtain
${d\over dt}M*\langle \phi,Au(t)\rangle= \langle\phi,u'(t)-\varphi(t)\rangle$, a.e.  $t\in (0,T)$, $\forall \phi\in X^*$ and since the right hand side of this equation vanishes for a.e. $t\in (t_0,T)$, the 
assertion $u=0$  directly follows from Theorem 1 of  \cite{KJM} 
and the injectivity of $A$.  This by \eqref{dpeq1} implies $\varphi=0$, too. In our case  have to provide a bit longer proof,  that however uses ideas of \cite{KJM}.

\medskip
\noindent{\it Proof}.
By \eqref{dpeq1} we have
\beqst
u(t)-AM*u(t)=1*\varphi(t)+{\rm y},\; t\in [0,T],
\eeqst
with some ${\rm y}\in X$. 
Using the assumptions  $\varphi(t)=0$ a.e. $t\in (t_0,T)$ and $u(t)=0$, $t\in (t_0,T)$ we obtain
\beq\label{th1p10}
AM*u(t)=-\int_0^{t_0}\varphi(\tau)d\tau-{\rm y},\quad t\in [t_0,T].
\eeq
Let us extend $u(t)$ by zero for $t>T$ and define $$F(t)=M*u(t),\;\;t\ge 0.$$ By $u(t)=0$, $t>t_0$,  we have
\beq\label{th1p9} 
F(t)=\int_0^{t_0} M(t-\tau)u(\tau)d\tau,\quad t>t_0.
\eeq
In view of  Bernstein's theorem, the completely monotonic function $M(t)$ has an analytic extension to $\Sigma(0,{\pi\over 2})$. 
Therefore, by Lemma \ref{lemanal} we deduce that 
 $F(t)$ is analytic for $t\in \Sigma(t_0,{\pi\over 2})$.
On the other hand, from \eqref{th1p10} we have $AF(t)={\rm x}$, $ t\in [t_0,T]$, where ${\rm x}=-\int_0^{t_0}\varphi(\tau)d\tau-{\rm y}$ is a constant element. 
Thus, $F(t)=A^{-1}{\rm x}$,  $ t\in [t_0,T]$.  Using analytic continuation we obtain 
\beq\label{th1p8}
F(t)=A^{-1}{\rm x}, \quad t\ge t_0.
\eeq
Next we express $F$ as follows:
$$
F(t)=F_1(t)+A^{-1}{\rm x},\quad t\ge 0.
$$
Due to \eqref{th1p8}, $F_1$ is compactly supported. 
Let $\phi$ be an arbitrary element of $X^*$ and apply the Laplace transform to the relation $\langle \phi,M*u\rangle=\langle \phi,F_1\rangle+\langle \phi,A^{-1}{\rm x}\rangle$. We obtain
$$
\widehat M(s)\langle\phi,\widehat u(s)\rangle=\langle\phi,\widehat F_1(s)\rangle+ {\langle\phi, A^{-1}{\rm x}\rangle\over s},\quad {\rm Re}s>0.
$$
If $u$ is not a zero function, then there exists $\tilde \phi\in X^*$ such that 
$\langle\tilde\phi,\widehat u(s)\rangle$ is not a zero function and we get 
\beq\label{th1p11}
\widehat M(s)={\langle\tilde\phi,\widehat F_1(s)\rangle+ {\langle \tilde\phi,A^{-1}{\rm x}\rangle\over s}\over \langle\tilde \phi,\widehat u(s)\rangle},\quad {\rm Re}s>0\, :
\, \langle\tilde\phi,\widehat u(s)\rangle\ne 0.
\eeq
Since $u$ and $F_1$ are compactly supported, $\langle\tilde\phi,\widehat u(s)\rangle$ and $\langle\tilde\phi,\widehat F_1(s)\rangle$ are entire functions.
This implies that the right-hand side of  
\eqref{th1p11}
can be meromorphically extended to the whole complex plane. But due to  $M\in \mathcal{CM}_\dagger$, this is not the case for the left-hand side. We reached the contradiction. Consequently, 
$u=0$. This with \eqref{dpeq1} implies $\varphi=0$.
\hfill $\Box$

\bigskip

\noindent{\it Proof of Theorem} \ref{th1}. Lemma \ref{thm1l1} implies $D(A_1)=D(A_2)$ and $\exists c>0$: $cM_1=M_2$, $A_1=cA_2$. Denote $M=M_1$, $A=A_1$, $u=u_1-u_2$ and $\varphi=f_1-f_2$. 
Then $M$, $A$, $u$ and $\varphi$ satisfy the assumptions of Lemma \ref{thm1l2}. Consequently, $u_1-u_2=0$, $f_1-f_2=0$. \hfill $\Box$

\medskip
We mention that input functions similar to $f$ in \eqref{th1v1} have been used earlier in other inverse problems to recover parameters of equations, e.g. in problems to determine 
coefficients  \cite{Kian2} and unknown manifolds \cite{Yl}.

\subsection{Applications}\label{Ex}

In this subsection  we will consider some inverse problems that are posed for equations governing subdiffusion in a 
\beq\label{domain}
\begin{split}
&\mbox{bounded domain $\Omega\subset \R^d$, $d\ge 1$, with a boundary $\partial\Omega$ }
\\
&\mbox{that of the class $C^2$ in case $d>1$. }
\end{split}
\eeq

Firstly, let us  consider inverse problems for equations with operators of the form
\eqref{Abasic}, 
where $\mathbf{a}=\big((a_{lm})|_{l,m=1,\ldots,d},(a_l)|_{l=1,\ldots,d},a\big)\in {\mathcal Q}$, 
\beqst
&&{\mathcal Q}=\Big\{\mathbf{a}\in (C(\overline\Omega))^{d^2+d+1}\, :\, a_{lm}=a_{ml}, \, l,m=1,\ldots,d,
\\
&&\qquad\mbox{$\exists \epsilon>0$: $\sum\limits_{l,m=1}^d a_{lm}(x)\xi_l\xi_m\ge\epsilon |\xi|^2$, $x\in\overline\Omega$, $\xi\in\R^d$}\Big\}.
\eeqst
Let us define 
\beq\label{dp}
D_p:=\{z\in W_p^2(\Omega)\, :\, z|_{\partial\Omega}=0\}, \quad 1<p<\infty,
\eeq
and an operator $A[p;\mathbf{a}]\,:\, D(A[p;\mathbf{a}])=D_p\mapsto L_p(\Omega)$ such that  $$A[p;\mathbf{a}]v=\mathscr{A}[\mathbf{a}]v\mbox{ for $v\in D_p$}.$$
The operator $A[p;\mathbf{a}]$ closed and densely defined in $L_p(\Omega)$. Moreover, it belongs to $\mathcal{S}(\omega,\theta)$ for some 
$\omega\in\R$, $\theta\in (0,{\pi\over 2}]$ \cite{Lun}.

\begin{corollary}\label{coro1} 
 Let $M_1\in \mathcal {CM}_\dagger$, $M_2\in \mathcal{CM}$,  
 $\mathbf{a}_j\in {\mathcal Q}$, $j=1,2$, $p\in (1,\infty)$,  $0\not\in \sigma(A[p,\mathbf{a}_1])$. Let $f_j$, $u_j$, $j=1,2$, satisfy 
  $ f_j\in L_1((0,T);L_p(\Omega))$, $u_j\in C([0,T];L_p(\Omega))$,  $M_j*u_j\in C([0,T];D_p)$,  
 $u_j-A[p;\mathbf{a}_j]M_j*u_j\in W_1^1((0,T);L_p(\Omega))$ and solve the equations
\beq\label{coreq1}
{\partial\over \partial t}\left[u_j(t,x)-A[p;\mathbf{a}]M_j*u_j(t,x)\right]=f_j(t,x),\; a.e.\; (t,x)\in (0,T)\times \Omega.
\eeq
Moreover, assume that $f_1(t,\cdot)=f_2(t,\cdot)=:f(t,\cdot)$ a.e. $t\in (t_0,T)$,  for some $t_0\in (0,T)$, $f(t,\cdot)=0$, a.e. $t\in (t_0,t_1)$ 
for some $t_1\in (t_0,T)$ and 
\beq\label{cor11}\left.
\begin{split}
&f(t,x)=\sum_{i=1}^\infty \psi_i(t)w_i(x),\; t\in (t_1,T),\; \psi_i\in L_1(t_1,T),\;
\\
&{\rm ess\, supp}\,\psi_i\subset [t_{i},t_{i+1}),\;  t_1<t_2<\ldots <T,\; 
\\
&t_{i}=\min{\rm ess\, supp}\,\psi_i,\;  
\;
{\rm span}\{w_i\, :\, i\in\N\}  - \mbox{dense in  $L_p(\Omega)$}.
\end{split}\right\}
\eeq
Finally, let $u_1(t,\cdot)=u_2(t,\cdot)$,  $t\in (t_0,T)$.  Then there exists $c>0$ such that $cM_1=M_2$, $\mathbf{a}_1=c\mathbf{a}_2$,
$f_1=f_2$, and $u_1=u_2$.
\end{corollary}

\noindent{\it Proof}. 
Theorem \ref{th1} implies that  
$f_1=f_2$, $u_1=u_2$ and
there exists $c>0$ such that $cM_1=M_2$ and $A[p;\mathbf{a}_1]=cA[p;\mathbf{a}_2]$. Let us 
denote the components of $\mathbf{a}_j$, $j=1,2$,  as follows: 
$$\mathbf{a}_j=\big((a_{lm,j})|_{l,m=1,\ldots,d},(a_{l,j})|_{l=1,\ldots,d},a_{,j}\big).$$
Let us choose some $x^*\in\Omega$ and $v^*\in D_p$ such that $v^*(x)=1$ for $x$ in some neighborhood of $x^*$. Then $A[p;\mathbf{a}_j]v^*(x)|_{x=x^*}=a_{,j}(x^*)$. Thus, 
$a_{,1}(x^*)=ca_{,2}(x^*)$.  Secondly, let $v^l(x)=v^*(x)(x_l-x_l^*)$, $l=1,\ldots,d$. Then $A[p;\mathbf{a}_j]v^l(x)|_{x=x^*}=a_{l,j}(x^*)$.
Thus, $a_{l,1}(x^*)=ca_{l,2}(x^*)$, $l=1,\ldots,d$.  Finally, working similarly with $v^{lm}(x)=v^*(x)(x_l-x_l^*)(x_m-x_m^*)$ we deduce
$a_{lm,1}(x^*)=ca_{lm,2}(x^*)$, $l,m=1,\ldots,d$. Since $x^*\in\Omega$ is arbitrary  we obtain 
$\mathbf{a}_1=c\mathbf{a}_2$. \hfill $\Box$

\bigskip
Secondly, we consider problems with distributed fractional elliptic operators. To this end we firstly introduce an elliptic operator $B_a$
by the formula
$$
B_av(x)=\Delta v(x)+a(x)v(x),\quad x\in\Omega,
$$
where $a\in C(\overline\Omega)$. It maps the set $D_2$ defined by \eqref{dp} to $L_2(\Omega)$. Assume that $a\le 0$ and introduce the eigenvalues and orthonormal in $L_2(\Omega)$ eigenfunctions of $-B_a$:
\beq\label{bqeigen}
\begin{split}
&-\Delta v_k(x)-a(x)v_k(x)=\mu_k v_k(x),\; x\in\Omega,\; v_k|_{\partial\Omega}=0,\quad k\in\N,
\\
&\qquad 0<\mu_1\le\mu_2\le\ldots,\;\; \langle v_k,v_l\rangle=\delta_{kl},
\end{split}
\eeq
where $\langle\cdot,\cdot\rangle$ is the scalar product in $L_2(\Omega)$. 
Next, let us define the distributed 
fractional powers of $B_a$ by the formula
\beqst
&&A_{\varrho,a}v:=-\int_0^1 (-B_a)^\beta d\varrho(\beta)v=-\sum_{k=1}^\infty \lambda_k \langle v,v_k\rangle v_k,
\\
&&\quad \lambda_k =\int_0^1 \mu_k^\beta d\varrho(\beta),
\eeqst
where $\varrho$ is a nonnegative Borel measure such that ${\rm supp}\,\varrho\cap (0,1]\ne\emptyset$.

The domain of $A_{\varrho,a}$ is 
$$
D(A_{\varrho,a})=\left\{v\in L_2(\Omega)\, :\, \sum_{k=1}^\infty \lambda_k^2 |\langle v,v_k\rangle|^2<\infty\right\}.
$$
Obviously, $\lambda_k$ and $v_k$ are the eigenvalues and eigenfunctions of $-A_{\varrho,a}$, i.e.
$$-A_{\varrho,a} v_k=\lambda_k v_k,\quad k\in\N,
$$
and $0<\lambda_1\le\lambda_2\le\ldots$. 

Since ${\rm span}\{v_k\, :\,k\in\N\}$ is dense in $L_2(\Omega)$  and forms a subset of $D(A_{\varrho,a})$,  $D(A_{\varrho,a})$ is dense in $L_2(\Omega)$. 
Further, 
$$
(\lambda I-A_{\varrho,a}) v=\sum_{k=1}^\infty (\lambda+\lambda_k) \langle v,v_k\rangle v_k.
$$
This shows that or any $\lambda\in \Sigma(0,\pi)$, the inverse  $(\lambda I-A_{\varrho,a})^{-1}$ exists, belongs to $\mathcal{B}(L_2(\Omega))$  and has the formula
$$
(\lambda I-A_{\varrho,a})^{-1} v=\sum_{k=1}^\infty {1\over \lambda+\lambda_k} \langle v,v_k\rangle v_k,\;\; \lambda\in \Sigma(0,\pi).
$$
Since $\rho(A_{\varrho,a})$ is not empty, $A_{\varrho,a}$ is closed. Moreover, for any $\theta'\in (0,\pi)$ and $\lambda \in \Sigma(0,\theta')$ we have
\beqst
&&|\lambda+\lambda_k|^2=|\lambda|^2+2 |\lambda|\lambda_k\cos\arg \lambda+\lambda_k^2\ge|\lambda|^2+ 2 |\lambda|\lambda_k\cos\theta'+\lambda_k^2
\\
&&=|\lambda|^2\sin^2\theta'+(|\lambda|\cos\theta'+\lambda_k)^2\ge |\lambda|^2\sin^2\theta'.
\eeqst
Hence
$$
\|(\lambda I-A_{\varrho,a})^{-1} v\|=\sqrt{\sum_{k=1}^\infty {1\over |\lambda+\lambda_k|^2} |\langle v,v_k\rangle|^2}\le {1\over|\lambda|\sin\theta'}\,\|v\|.
$$
Consequently, $A_{\varrho,a}\in \mathcal{S}(0,\theta')$ for any $\theta'\in (0,{\pi\over 2})$.

\begin{corollary}\label{coro2}
 Let $M_1\in \mathcal {CM}_\dagger$, $M_2\in \mathcal{CM}$,  $a_j\in C(\overline\Omega)$, $a_j\le 0$, $j=1,2$, 
 and measures $\varrho_j$, $j=1,2$, satisfy either
 \beq\label{cor21}\left.
 \begin{split}
&\varrho_j=\sum_{l=1}^{n_j}\varkappa_{l,j}\delta_{\beta_{l,j}},\; j=1,2,\; \quad\mbox{where}
\\
&n_j\in\N, \, \; 0<\beta_{1,j}<\beta_{2,j}<\ldots<\beta_{n_j,j}\le 1,\; \varkappa_{l,j}>0,
 \\
 &\mbox{$\delta_{\beta_{l,j}}$ - the point measure concentrated at $\beta_{l,j}$}
 \end{split}\right\}
 \eeq
 or
  \beq\label{cor22}
\left.   \begin{split}
&\mbox{$\varrho_j$ - absolutely continuous}, \; j=1,2,
\\
&\varrho_j(U)=\int_U\varkappa_j(\beta)d\beta,\; \;
\mbox{$U$ - any Borel subset of $(0,1)$},\;
\\
&\mbox{there exists $\delta>0$ such that $\varkappa_j$ is analytic in $(0 ,1+\delta)$}.
\end{split}\right\}
\eeq
Let $f_j$, $u_j$, $j=1,2$, satisfy $ f_j\in L_1((0,T);L_2(\Omega))$, $u_j\in C([0,T];L_2(\Omega))$,  $M_j*u_j\in C([0,T];X_{A_{\varrho_j,a_j}})$, 
 $u_j-A_{\varrho_j,a_j}M_j*u_j\in W_1^1((0,T);L_2(\Omega))$ and solve the equations
\beq\label{coreq2}
{\partial\over \partial t}\left[u_j(t,x)-A_{\varrho_j,a_j}M_j*u_j(t,x)\right]=f_j(t,x),\; a.e.\; (t,x)\in (0,T)\times \Omega.
\eeq
Moreover, assume that $f_1(t,\cdot)=f_2(t,\cdot)=:f(t,\cdot)$ a.e. $t\in (t_0,T)$,  for some $t_0\in (0,T)$, $f(t,\cdot)=0$, a.e. $t\in (t_0,t_1)$ 
for some $t_1\in (t_0,T)$ and
\beq\label{cor23}\left.
\begin{split}
&f(t,x)=\sum_{i=1}^\infty \psi_i(t)w_i(x),\; t\in (t_1,T),\; \psi_i\in L_1(t_1,T),\;
\\
&{\rm ess\, supp}\,\psi_i\subset [t_{i},t_{i+1}),\;  t_1<t_2<\ldots <T,\; 
\\
&t_{i}=\min{\rm ess\, supp}\,\psi_i,\;  
\;
{\rm span}\{w_i\, :\, i\in\N\}  - \mbox{dense in  $L_2(\Omega)$}.
\end{split}\right\}
\eeq
Finally, let $u_1(t,\cdot)=u_2(t,\cdot)$,  $t\in (t_0,T)$.  Then there exists $c>0$ such that $cM_1=M_2$, $\varrho_1=c\varrho_2$, 
$a_1=a_2$,
$f_1=f_2$ and $u_1=u_2$.
\end{corollary}

\noindent{\it Proof}. To apply Theorem \ref{th1} we have to verify  that $D(A_{\varrho_1,a_1})\cap D(A_{\varrho_2,a_2})$ is not empty and dense in $X_{A_{\varrho_j,a_j}}$, $j=1,2$. Other 
assumptions of this theorem are evident. 
Let $0<\mu_{1,j}\le\mu_{2,j}\le\ldots$ be the Dirichlet eigenvalues of  $-\Delta -a_jI$ in $\Omega$  and $v_{1,j}$, $v_{2,j}, \ldots$ the corresponding  orthonormal in $L_2(\Omega)$ eigenfunctions.
Let us denote $\lambda_{k,j}=\int_0^1 \mu_{k,j}^\beta d\varrho_j(\beta)$.
Then  for any $v\in D_2$, where $D_2$ is given by \eqref{dp}, we have
\beqst
&&
\|v\|_{X_{A_{\varrho_j,a_j}}}^2=\sum_{k=1}^\infty \lambda_{k,j}^2|\langle v,v_{k,j}\rangle|^2=\sum_{k=1}^\infty \left(\int_0^1 \mu_{1,j}^{\beta}\left[{\mu_{k,j}\over\mu_{1,j}}\right]^\beta d\varrho_j(\beta)\right)^2
\\
&&\times |\langle v,v_{k,j}\rangle|^2\le \sum_{k=1}^\infty \left(\int_0^1 \mu_{1,j}^{\beta} d\varrho_j(\beta){\mu_{k,j}\over\mu_{1,j}}\right)^2|\langle v,v_{k,j}\rangle|^2
\\
&&=\mu_{1,j}^{-2}\left(\int_0^1 \mu_{1,j}^{\beta} d\varrho_j(\beta)\right)^2\,\sum_{k=1}^\infty \mu_{k,j}^2|\langle v,v_{k,j}\rangle|^2,\; j=1,2.
\eeqst
The quantity  $\sqrt{\sum_{k=1}^\infty \mu_{k,j}^2|\langle v,v_{k,j}\rangle|^2}$ is a norm of $v$  in $D_2$. Therefore, 
$\|v\|_{X_{A_{\varrho_j,a_j}}}<\infty$ $j=1,2$. This proves $D_2\subset D(A_{\varrho_1,a_1})\cap D(A_{\varrho_2,a_2})$. In particular, ${\rm span}\, \{v_{k,j}, :\, k\in\N\}\subset D_2$, $j=1,2$. Next let
us choose some $j\in\{1;2\}$ and $v\in D(A_{\varrho_j,a_j})$. Then $\tilde v_n:=\sum_{k=1}^n \lambda_{k,j}\langle v,v_{k,j}\rangle v_{k,j}\in D_2$. 
It holds $\|v-\tilde v_n\|_{X_{A_{\varrho_j,a_j}}}\to 0$ as $n\to \infty$. This proves that $D_2$ is dense in  $X_{A_{\varrho_j,a_j}}$. Thus, 
$D(A_{\varrho_1,a_1})\cap D(A_{\varrho_2,a_2})$ is dense in $X_{A_{\varrho_j,a_j}}$, $j=1,2$.

From Theorem \ref{th1} we obtain that  
$f_1=f_2$,  $u_1=u_2$, $D(A_{\varrho_1,a_1})= D(A_{\varrho_2,a_2})$ and
there exists $c>0$ such that $cM_1=M_2$,  $A_{\varrho_1,a_1}=cA_{\varrho_2,a_2}$. 
Since $(\lambda_{k,j},v_{k,j})|_{k\in\N}$, are the pairs of eigenvalues and eigenfunctions of $-A_{\varrho_j,a_j}$, from the latter relation we have 
 $\lambda_{k,1}=c\lambda_{k,2}$, $v_{k,1}=v_{k,2}$, $k\in\N$. 
Therefore, denoting $v_k:=v_{k,1}=v_{k,2}$, we have 
\beq\label{cor2p1}-\Delta v_k -a_jv_k=\mu_{k,j}v_k,\quad k\in\N,\; j=1,2.
\eeq
Setting here $k=1$, subtracting the equations for $j=1$ and $j=2$ and dividing by $v_1(x)$ ($v_1$ preserves the sign \cite{Gil}, p. 214), we deduce
\beq\label{cor2p2}
a_1(x)=a_2(x)-\eta,\;\;x\in\Omega,\quad \eta:=\mu_{1,1}-\mu_{1,2}.
\eeq
Returning to \eqref{cor2p1} and using \eqref{cor2p2} we obtain
\beq\label{cor2p3}
\mu_{k,1}=\mu_{k,2}+\eta,\;\; k\in\N.
\eeq

Let us proceed to the relation $\lambda_{k,1}=c\lambda_{k,2}$. This by \eqref{cor2p3} implies 
\beq\label{cor2pr-1}
\int_0^1 (\mu_k+\eta)^\beta d\varrho_1(\beta)=c\int_0^1 \mu_k^\beta d\varrho_2(\beta),\;\;k\in\N,
\eeq
where $\mu_k=\mu_{k,2}$. We will analyze this equation separately in cases \eqref{cor21} and \eqref{cor22}. 

In case \eqref{cor21} we have 
\beq\label{cor2pr0}
\sum_{l=1}^{n_1}\varkappa_{l,1}(\mu_k+\eta)^{\beta_{l,1}}=c\sum_{l=1}^{n_2}\varkappa_{l,2}\mu_k^{\beta_{l,2}},\;\;k\in\N.
\eeq
Let us rewrite this in the following form:
\beq\label{cor2pr1}
&&\underbrace{\sum_{l=1}^{n_1}\varkappa_{l,1}\mu_k^{\beta_{l,1}}}_{W_{k,1}}\,-\,\underbrace{c\sum_{l=1}^{n_2}\varkappa_{l,2}\mu_k^{\beta_{l,2}}}_{W_{k,2}}=R_k,\quad k\in\N,
\\ \nonumber
&& R_k=\sum_{l=1}^{n_1}\varkappa_{l,1}\left[\mu_k^{\beta_{l,1}}-(\mu_k+\eta)^{\beta_{l,1}}\right].\;\; 
\eeq
Since  $\mu_k\to\infty$ as $k\to\infty$ and  $0<\beta_{1,j}<\beta_{2,j}<\ldots<\beta_{n_j,j}$, $j=1,2$, we have 
$W_{k,1}\sim \varkappa_{n_1,1}\mu_k^{\beta_{n_1,1}}\to \infty$ and $W_{k,2}\sim c\varkappa_{n_2,2}\mu_k^{\beta_{n_2,2}}\to \infty$ as $k\to\infty$. On the other hand, the sequence $R_k$ is bounded, because 
$$
\left|\mu_k^{\beta_{l,1}}-(\mu_k+\eta)^{\beta_{l,1}}\right|=\left|\int^{\mu_k}_{\mu_k+\eta}\beta_{l,1}\tau^{\beta_{l,1}-1}d\tau\right|\le \beta_{l,1}|\eta|
$$
for $k$ such that $\mu_k\ge 1$ and $\mu_k+\eta\ge 1$. The difference $W_{k,1}-W_{k,2}$ can be bounded in the process $k\to\infty$ only if $\beta_{n_1,1}=\beta_{n_2,2}$ and $\varkappa_{n_1,1}=c\varkappa_{n_2,2}$. Thus, from \eqref{cor2pr1} we obtain
\beqst
&&\underbrace{\sum_{l=1}^{n_1-1}\varkappa_{l,1}\mu_k^{\beta_{l,1}}}_{\tilde W_{k,1}}\,-\,\underbrace{c\sum_{l=1}^{n_2-1}\varkappa_{l,2}\mu_k^{\beta_{l,2}}}_{\tilde W_{k,2}}=R_k,\quad k\in\N.
\eeqst
Now $\tilde W_{k,1}\sim \varkappa_{n_1-1,1}\mu_k^{\beta_{n_1-1,1}}\to \infty$ and $\tilde W_{k,2}\sim c\varkappa_{n_2-1,2}\mu_k^{\beta_{n_2-1,2}}\to \infty$ as $k\to\infty$. Using again the 
fact that $R_k$ is bounded we deduce  $\beta_{n_1-1,1}=\beta_{n_2-1,2}$ and $\varkappa_{n_1-1,1}=c\varkappa_{n_2-1,2}$. Continuing such arguments we prove that 
\beq\label{cor2pr2}
n_1=n_2=:n,\;\; \varkappa_{l,1}=c\varkappa_{l,2},\; \beta_{l,1}=\beta_{l,2},\; l=1,\ldots n.
\eeq
This by \eqref{cor21} implies $\varrho_1=c\varrho_2$. 
From \eqref{cor2pr0}  by means of  \eqref{cor2pr2} we have
\beq\label{cor2p9}
\sum_{l=1}^{n}\varkappa_{l,1}(\mu_1+\eta)^{\beta_{l,1}}=\sum_{l=1}^{n}\varkappa_{l,1}\mu_1^{\beta_{l,1}}.
\eeq
The function $\Psi(\xi)=\sum_{l=1}^{n}\varkappa_{l,1}(\mu_1+\xi)^{\beta_{l,1}}$ is strictly increasing in $(-\mu_1,\infty)$. Therefore, \eqref{cor2p9} implies $\eta=0$. From 
\eqref{cor2p2} we obtain $a_1=a_2$. 

Secondly, we consider the case  \eqref{cor22}. Then from \eqref{cor2pr-1} we have
\beq\label{cor2pr10}
\int_0^1 \varkappa_1(\beta)(\mu_k+\eta)^\beta d\beta=c\int_0^1 \varkappa_2(\beta)\mu_k^\beta d\beta,\;\;k\in\N.
\eeq
This yields
\beq\label{cor2pr11}
&&\int_0^1[ \varkappa_1(\beta)-c\varkappa_2(\beta)]\mu_k^\beta d\beta=\overline R_k,\;\;k\in\N,
\\ \nonumber 
&&\overline R_k=\int_0^1 \varkappa_1(\beta)[\mu_k^\beta -(\mu_k+\eta)^\beta]d\beta.
\eeq
Like in the previous case, we can show that the sequence $\overline R_k$ is bounded. Let us choose an arbitrary $\beta_*\in (0,1)$ and suppose that $\varkappa_1(\beta)-c\varkappa_2(\beta)$ differs from 
zero and preserves the sign in the interval $(\beta_*,1)$. Then there exists $\beta_{**}\in (\beta_*,1)$ and $\epsilon\in (0,\beta_{**}-\beta_{*})$ such that $|\varkappa_1(\beta)-c\varkappa_2(\beta)|>\epsilon$, $|\beta-\beta_{**}|<\epsilon$. 
Thus for $k$ such that $\mu_k\ge 1$ we obtain
\beqst
&&\left|\int_0^1[ \varkappa_1(\beta)-c\varkappa_2(\beta)]\mu_k^\beta d\beta\right|\ge\left|\int_{\beta_{*}}^1[ \varkappa_1(\beta)-c\varkappa_2(\beta)]\mu_k^\beta d\beta\right| 
\\
&&-
\left|\int_0^{\beta_*}[ \varkappa_1(\beta)-c\varkappa_2(\beta)]\mu_k^\beta d\beta\right|\ge \int_{\beta_{**}-\epsilon}^{\beta_{**}+\epsilon}| \varkappa_1(\beta)-c\varkappa_2(\beta)|\mu_k^\beta d\beta
\\
&&-
\left|\int_0^{\beta_*}[ \varkappa_1(\beta)-c\varkappa_2(\beta)]\mu_k^\beta d\beta\right|\ge 2\epsilon^2\mu_k^{\beta_{**}-\epsilon}-
C_*\mu_k^{\beta_{*}},
\eeqst
where $C_*=\int_0^{\beta_*}| \varkappa_1(\beta)-c\varkappa_2(\beta)|d\beta$. Since $\beta_{**}-\epsilon>\beta_{*}$, we have \break
$\left|\int_0^1[ \varkappa_1(\beta)-c\varkappa_2(\beta)]\mu_k^\beta d\beta\right|\to\infty$
as $k\to\infty$. This is in the contradiction with the fact that the right-hand side of \eqref{cor2pr11} is bounded. Therefore,  $ \varkappa_1(\beta)-c\varkappa_2(\beta)$ cannot preserve the sign in the 
interval $(\beta_*,1)$. Since $\beta_*\in (0,1)$ is arbitrary, the zeros of $ \varkappa_1(\beta)-c\varkappa_2(\beta)$ have the accumulation point $\beta=1$. The functions $\varkappa_j(\beta)$ are analytic 
in $(0,1+\delta)$, therefore we obtain $\varkappa_1(\beta)=c\varkappa_2(\beta)$, $\beta\in (0,1+\delta)$. This implies $\varrho_1=c\varrho_2$. From \eqref{cor2pr10} we deduce 
$$
\int_0^1 \varkappa_1(\beta)(\mu_1+\eta)^\beta d\beta=\int_0^1 \varkappa_1(\beta)\mu_1^\beta d\beta.
$$
The function $\Phi(\xi)=\int_0^1 \varkappa_1(\beta)(\mu_1+\xi)^\beta d\beta$ is strictly increasing. Therefore, $\eta=0$ and from \eqref{cor2p2} we obtain $a_1=a_2$. 
\hfill $\Box$

\bigskip
If a priori $a_1=a_2$   then the eigenfunctions of $A_{\varrho_1,a_1}$ and $A_{\varrho_2,a_2}$  coincide and we can formulate a  uniqueness result also in case  $f$ satisfies \eqref{th1v2}. 
Let us formulate such a statement without a proof.

\begin{corollary}\label{coro3}
Let $M_1\in \mathcal {CM}_\dagger$, $M_2\in \mathcal{CM}$, $a\le 0$,
 and either \eqref{cor21} or \eqref{cor22} hold.
 Let $f_j$, $u_j$, $j=1,2$, satisfy $ f_j\in L_1((0,T);L_2(\Omega))$,  $u_j\in C([0,T];L_2(\Omega))$,  $M_j*u_j\in C([0,T];X_{A_{\varrho_j,a}})$, 
 $u_j-A_{\varrho_j,a}M_j*u_j\in W_1^1((0,T);L_2(\Omega))$ and solve 
\beq\label{coreq3}
{\partial\over \partial t}\left[u_j(t,x)-A_{\varrho_j,a}M_j*u_j(t,x)\right]=f_j(t,x),\; a.e.\; (t,x)\in (0,T)\times \Omega.
\eeq
Moreover, assume that $f_1(t,\cdot)=f_2(t,\cdot)=:f(t,\cdot)$ a.e. $t\in (t_0,T)$,  for some $t_0\in (0,T)$, $f(t,\cdot)=0$, a.e. $t\in (t_0,t_1)$ 
for some $t_1\in (t_0,T)$ and $f$ satisfies either \eqref{cor23} or 
\beqst\label{cor31}
\begin{split}
&\mbox{$\langle f(t),v_k\rangle\not\equiv 0$ in $(t_1,T)$,  $k\in\N$, where $v_k$ are the Dirichlet}
\\
&\mbox{eigenfunctions of $\Delta+aI$ (i.e. solutions of \eqref{bqeigen}).}
\end{split}
\eeqst
Finally, let $u_1(t,\cdot)=u_2(t,\cdot)$,  $t\in (t_0,T)$.  Then there exists $c>0$ such that $cM_1=M_2$, $\varrho_1=c\varrho_2$, 
$f_1=f_2$ and $u_1=u_2$.
\end{corollary}

\section{Inverse problems to determine the kernel}\label{sec:ker}

\subsection{Inverse problem to determine the kernel, source and state}\label{ss1}

Now we consider a problem to identify $M$, $u$ and $f$ using prescribed values of $u$ and $f$ in the interval $(t_0,T)$ in case $A$ is known. Then we are able prove the uniqueness in case of
 less restrictions on $f$. 

\begin{theorem}\label{thM1}
Let $M_1\in \mathcal {CM}_\dagger$,  $M_2\in \mathcal {CM}$, $A\, :\, D(A)\mapsto X$-  linear closed densely defined injective operator in a complex Banach space $X$,  $A\in \mathcal{S}(\omega,\theta)$
 for some 
 $\omega\in\R$, $\theta\in (0,{\pi\over 2}]$. Let $f_j$, $u_j$, $j=1,2$,  satisfy the assumptions \eqref{th1a1} - \eqref{th1a4} with $A_1=A_2=A$ and $f$ involved in \eqref{th1a4} fulfill the condition \eqref{th1a5}. 
Assume that $f(t)\not\equiv 0$ in $(t_1,T)$. 
Finally, let \eqref{th1a6} hold. 
 Then  $M_1=M_2$,  $f_1=f_2$ and $u_1=u_2$.
\end{theorem}

\noindent{\it Proof}. 
Applying Lemma  \ref{prop1} in case $A_1=A_2=A$ , we have\break  $\int_{t_1}^tS_1(t-\tau)f(\tau)d\tau=\int_{t_1}^tS_2(t-\tau)f(\tau)d\tau$, $t\in [t_1,T]$, where $S_j$ is the resolvent of \eqref{ab1} with $M=M_j$. Let us extend $f(t)$ by $0$ for $t<t_0$. Then we obtain $w_1(t)=w_2(t)$, $t\in [0,T]$, where $w_j(t)=S_j*f(t)$. 
Due to Lemma \ref{pruss} (ii), the function  $w_j$ satisfies the equation $w_j(t)-AM_j*w_j(t)=1*f(t)$, $t\in [0,T]$. Therefore, denoting $w:=w_1=w_2$, we obtain 
$A(M_1-M_2)*w(t)=0$, $t\in [0,T]$. Since $A$ is injective, we have 
\beq\label{thM1p1}
(M_1-M_2)*w(t)=0,\;\;t\in [0,T].
\eeq
 The function $w(t)$ is not identically zero, because otherwise the left-hand side of  the equation 
$w(t)-AM_j*w(t)=1*f(t)$ vanishes and this contradicts the assumption $f\not\equiv 0$. Therefore, there exists $\phi\in X^*$ such that $\langle\phi, w(t)\rangle\not\equiv 0$. 
Denote $t_w=\min\, {\rm ess\, supp} \langle\phi, w\rangle$. Then from \eqref{thM1p1} we deduce
\beqst
\int_{t_w}^t \langle\phi, w(\tau)\rangle (M_1(t-\tau)-M_2(t-\tau))d\tau=0,\;\; t\in [t_w,T].
\eeqst
The function $\langle\phi, w(t)\rangle$ does not vanish identically in an arbitrarily small right neighborhood of $t_w$. Titchmarsh convolution theorem implies 
$M_1(t)=M_2(t)$, $t\in (0,T-t_w]$ and by analytic continuation we obtain $M_1=M_2$.

The assertions $f_1=f_2$ and $u_1=u_2$ can be deduced by means of Lemma \ref{thm1l2} like in the proof of Theorem \ref{th1}.
\hfill $\Box$

\medskip Theorem \ref{thM1} applies to inverse problems for generalized fractional diffusion equations with operators defined in Subsection \ref{Ex}. Let 
$\Omega$ meet the requirements \eqref{domain} and either
\beqst
&&\hskip-5truemm X=L_p(\Omega)\; \mbox{for some $p\in (1,\infty)$},\;\; {\mathbf a}\in \mathcal{Q}, \;\; 
A=A[p;\mathbf{a}],\; 0\not\in \sigma(A[p;\mathbf{a}])
\eeqst
or
\beqst
&&\hskip-5truemm X=L_2(\Omega),\; \mbox{$\varrho$ - a nonnegative Borel measure},\; 
\\
&&a\in C(\overline\Omega), \;a\le 0,
\;A=A_{\varrho,a}.
\eeqst
In both cases $A$ satisfies the assumptions of Theorem \ref{thM1}.

\subsection{Inverse problem to determine the kernel in case of unknown history}\label{ss2}

In this section we consider the case when a functional of  $u(t)$ is given after $t=t_0$. This information may not suffice to reconstruct the history of 
$f$ and $u$ before $t=t_0$ but it still enables to recover the kernel $M$. 

Firstly let us introduce the square of $A$ and formulate a lemma. Let $A\, :\, D(A)\mapsto X$ -  linear closed densely defined operator in a complex Banach space $X$,
$A\in \mathcal{S}(\omega,\theta)$ for some $\omega\in\R$, $\theta\in (0,\pi/2]$. From the fact that $\rho(A)$ is not empty, we can deduce that $D(A^2):=\{{\rm x}\in D(A)\, :\, A{\rm x}\in D(A)\}\ne\emptyset$.
Let us denote by $X_{A^2}$ the set 
$D(A^2)$ endowed with the norm 
\beq\label{Phinorm}
\|{\rm x}\|_{X_{A^2}}=\|{\rm x}\|+\|A{\rm x}\|+\|A^2{\rm x}\|. 
\eeq
The closedness of $A$ implies that $X_{A^2}$ is a Banach space with respect to the norm \eqref{Phinorm}.

\begin{lemma}\label{prop2}
Let $M_j\in \mathcal {CM}$, $j=1,2$,  $A\, :\, D(A)\mapsto X$ -  linear closed densely defined operator in a complex Banach space $X$,  $A\in \mathcal{S}(0,\theta)$
 for some 
 $\theta\in (0,{\pi\over 2}]$,  $g\in W^{1,1}((0,T);X_{A^{2}})$ and  $w_j$, $j=1,2$, satisfy $w_j\in C([0,T];X_{A^{2}})$ and
  solve 
$$w_j(t)-M_j*Aw_j(t)=g(t),\;\;t\in [0,T].$$
Moreover, let $\Phi\in X^*$,  $\langle\Phi,Ag(0)\rangle\ne 0$ and $\langle\Phi,w_1(t)\rangle=\langle\Phi,w_2(t)\rangle$, $t\in (0,T)$.  Then  $M_1=M_2$.
\end{lemma}

\noindent {\it Proof}. In Theorem 5.1 of \cite{JKas}, the equality $M_1(t)=M_2(t)$, $t\in (0,T)$, was proved in case $A$, $w_j$, $g$ satisfy the assumptions of Lemma \ref{prop2} but $M_j$, $j=1,2$,  belong to a class  
of kernels that is more general than $\mathcal {CM}$. Since we assume $M_j\in \mathcal {CM}$, the relation $M_1(t)=M_2(t)$, $t>0$, follows by analytic continuation. \hfill $\Box$

\begin{theorem}\label{thmviim}
Let $M_j\in \mathcal {CM}$, $j=1,2$,  $A\, :\, D(A)\mapsto X$ -  linear closed densely defined operator in a complex Banach space $X$,  $A\in \mathcal{S}(0,\theta)$
 for some 
 $\theta\in (0,{\pi\over 2}]$ and $\Phi\in X^*$. Let $f_j$, $u_j$, $j=1,2$,  satisfy the assumptions \eqref{th1a1} - \eqref{th1a4} with $A_1=A_2=A$ and $f$ involved in \eqref{th1a4} fulfill the condition \eqref{th1a5}.
 Moreover, assume that
  \beq\label{th3v1}
  \begin{split}
 & \exists m\in\{0\}\cup \N\, :\, f|_{(t_1,T)}\in W_1^{m+1}((t_1,T);X_{A^{2}}),
 \\
 &\langle\Phi, f^{(j)}(t_1^+)\rangle=0,\, j=0,\ldots,m-1\;\;  \mbox{(if $m\ge 1$),}\;\; 
 \langle\Phi,A f^{(m)}(t_1^+)\rangle\ne 0.
  \end{split}
  \eeq
Finally, let  $\langle\Phi,u_1(t)\rangle=\langle\Phi,u_2(t)\rangle$, $t\in (t_0,T)$.  Then  $M_1=M_2$.
\end{theorem}

\noindent{\it Proof}. 
We have 
\beq\label{th2p1}
u_j(t)=S_j(t)u_j(0)+S_j*f_j(t),\quad t\in [0,T],\;\; j=1,2,
\eeq
where $S_j$ is the resolvent of the equation \eqref{ab1} with $M=M_j$. 
Hence,
\beqst
\langle\Phi,u_j(t)\rangle=\langle\Phi,S_j(t)u_j(0)\rangle+\langle\Phi,S_j*f_j(t)\rangle,\quad t\in [0,T],\;\; j=1,2.
\eeqst
Let $t\in [t_0,T]$.  Subtracting this relation with $j=2$ from the one with $j=1$ and using the assumptions 
$\langle\Phi,u_1(t)\rangle=\langle\Phi,u_2(t)\rangle$, $t\in (t_0,T)$, and $f_1(t)=f_2(t)=f(t)$, $t\in (t_0,T)$, we deduce 
\beqst
\tilde Y(t)+\tilde Z(t)=0,\quad t\in [t_0,T],
\eeqst
where 
\beqst
&&\tilde Y(t)=\Big\langle\Phi,S_1(t)u_1(0)-S_2(t)u_2(0)
\\
&&\qquad +\int_0^{t_0}S_1(t-\tau)f_1(\tau)d\tau-\int_0^{t_0}S_2(t-\tau)f_2(\tau)d\tau\Big\rangle,
\\
&&\tilde Z(t)=\Big\langle\Phi,\int_{t_0}^t (S_1(t-\tau)-S_2(t-\tau))f(\tau)d\tau\Big\rangle.
\eeqst
The function $\tilde Y(t)$ is analytic in $(t_0,T)$ and $\tilde Z(t)$ vanishes in $(t_0,t_1)$ because of the assumption $f(t)=0$, $t\in (t_0,t_1)$. Arguing as in the proof of 
Lemma \ref{prop1}, we obtain 
$\tilde Z(t)=0$, $t\in [t_0,T]$. Thus in view of $f|_{(t_0,t_1)}=0$ we have
\beqst
&&\Big\langle\Phi,\int_{0}^t S_1(\tau)f(t_1+t-\tau)d\tau\Big\rangle=\Big\langle\Phi,\int_{0}^t S_2(\tau)f(t_1+t-\tau)d\tau\Big\rangle,\;\;
\\
&&\qquad t\in [0,T-t_1].
\eeqst
Differentiating this relation $m$ times and observing the assumptions $\langle\Phi, f^{(j)}(t_1^+)\rangle=0$, $j=0,\ldots,m-1$, we obtain $\langle\Phi,w_1(t)\rangle=\langle\Phi,w_2(t)\rangle$, $t\in [0,T-t_1]$, where $$w_j(t)=S_j(t)g(0)+\int_0^t S_j(t-\tau)g'(\tau)d\tau$$ and $g(t)=f^{(m)}(t_1+t)$. 
By $f|_{(t_1,T)}\in W_1^{m+1}((t_1,T);X_{A^{2}})$ we have  $g\in W_1^1 ((0,T-t_1),X_{A^2})$.  This implies that $w_j$ belongs to $C([0,T-t_1], X_{A^2})$. Due to Lemma \ref{pruss} (iii),  $w_j$ solves $w_j(t)-M_j*Aw_j(t)=g(t)$, $t\in [0,T-t_1]$. 
Moreover, by $\langle\Phi,A f^{(m)}(t_1^+)\rangle\ne 0$ we have $\langle\Phi,Ag(0)\rangle\ne 0$. Assumptions of 
Lemma \ref{prop2} are satisfied. We obtain $M_1=M_2$. \hfill $\Box$

\bigskip
If  $X$ is a space of functions defined in $\Omega$, then the functional $\Phi$  may represent for instance an integral observation of the form 
$\langle\Phi,v\rangle=\int_{\Omega'} v(x)dx$, $\Omega'\subseteq\Omega$ (then we can work in $L_p$ spaces as in the previous examples) or a point observation $\langle\Phi,v\rangle=v(x_0)$, 
$x_0\in\overline\Omega$ (then we have to work in a space of continuous functions). Let us consider the case of point observation and involve also Neumann boundary conditions. The latter option
enables to observe on the boundary, too. 

Assume that $\Omega$ meets the requirements \eqref{domain} and either
\beqst
&&X=C_0(\overline\Omega)=\{v\in C(\overline\Omega)\, :\, v|_{\partial\Omega}=0\}, \;
\\
&&D(A)=\Big\{v\in C_0(\overline\Omega)\,:\,
v\in \bigcap\limits_{p>1}W_p^{2}(\Omega), \,   \Delta v\in C_0(\overline\Omega)\Big\}, \quad 
\\
&&Av=\Delta v,\, v\in D(A),\quad \langle\Phi,v\rangle =v(x_0),\, x_0\in \Omega
\eeqst
or 
\beqst
&&X=C(\overline\Omega), 
\\
&&D(A)=\Big\{v\in C(\overline\Omega)\,:\,v\in \bigcap\limits_{p>1}W_p^{2}(\Omega), \,  {\partial\over\partial\nu}v|_{\partial\Omega}=0,\,  \Delta v\in C(\overline\Omega)\Big\},
\\
&&\mbox{$\nu$ -  the outer normal of $\partial\Omega$}, 
\\
&&Av=\Delta v,\, v\in D(A),\quad \langle\Phi,v\rangle =v(x_0),\, x_0\in \overline\Omega.
\eeqst
In both these cases $A$ and $\Phi$ satisfy the assumptions  of Theorem \ref{thmviim} (for $A$, see e.g. \cite{Pruss}, p. 139). 

\section{Concluding remarks}\label{conc}

The assumption $f|_{(t_0,t_1)}=0$ is essential in the proof of Lemma \ref{prop1}.  If $f$ does not vanish identically in some right neighborhood of $t_0$, then we cannot exclude $Y(t)$ 
from \eqref{th1p2} and derive an independent equation for $M_j$ and $A_j$. 

Let us give an interpretation for \eqref{th1p4e}. To this end define the functions $\tilde u_j\, :\, (0,T-t_1)\mapsto X$  by
 $\tilde u_j(t-t_1)=\int_{t_1}^t S_j(t-\tau)f(\tau)d\tau$, $t\in (t_1,T)$. They satisfy the following equations:
$$
{d\over dt}\left[\tilde u_j(t)-A_jM_j*\tilde u_j(t)\right]=f(t+t_1),\; \; a.e.\;\; t\in (0,T-t_1).
$$
Then \eqref{th1p4e} implies that the solutions of these equations concide. In other words, in Lemma \ref{thm1l1} we proved the uniqueness for an inverse problem to recover $MA$  in the 
 equation   ${d\over dt}\left[\tilde u(t)-AM*\tilde u(t)\right]=f(t_1+t)$, $t\in (0,T-t_1)$, by means of fully given $\tilde u$. 

Theorem \ref{th1} was proved in two stages: firstly we proved
 $M_1A_1=M_2A_2$ and thereupon $f_1=f_2$ and $u_1=u_2$.
Similar decomposition of an inverse problem into two stages was used in Theorem \ref{thM1}, too. Firstly, the uniqueness for $M$ and then the uniqueness for  source and state was established there.

The function $f$ satisfying conditions of  Theorems \ref{th1}, \ref{thM1} is not analytic in $(t_0,T)$. A singularity occurs in $[t_1,T)$.  This kind of irregularity characterizes our method.

Theorem \ref{thmviim} also uses  a decomposition of the inverse problem into two stages, but the uniqueness was shown only in the first stage, i.e. for $M$. 
The history of $f$ and $u$ remains unknown. We believe that a similar incorporation of unknown history may be possible also in other inverse problems 
to recover parameters of fractional diffusion equations, e.g. problems to determine coefficients by means of boundary data.

\bigskip\noindent
{\bf Acknowledgement}. \\[1ex] The study was supported by Estonian Research Council Grant PRG832.

\end{document}